\documentclass[
amsmath,amssymb,
aps, nofootinbib
rmp, floatfix, reprint,
superscriptaddress,
groupedaddress, showkeys
]{revtex4-2}
\usepackage{graphicx}
\usepackage{dcolumn}
\usepackage{bm}

\usepackage[toc,page,header]{appendix}
\makeatletter
\newcommand\appendix@section[1]{%
  \refstepcounter{section}%
  \orig@section*{Appendix \@Alph\c@section: #1}%
}
\let\orig@section\section
\g@addto@macro\appendix{\let\section\appendix@section}
\makeatother

\usepackage{amsmath,amssymb}
\usepackage{slashed}
\usepackage{graphicx}
\usepackage{dsfont}
\usepackage{bm}
\usepackage[top=1.0in,bottom=1.0in,left=1.0in,right=1.0in]{geometry}
\usepackage[colorlinks,linkcolor=blue,citecolor=blue]{hyperref}
\usepackage{CJKutf8}
\usepackage{tikz-feynhand}
\usepackage[caption=false]{subfig}
\usepackage{float}

\begin{document}
\setlength{\oddsidemargin}{0.3cm}
\setlength{\topmargin}{-0.1cm}
\setlength{\textheight}{21cm}
\setlength{\textwidth}{16cm}
\newcommand{\be}{\begin{equation}}
\newcommand{\ee}{\end{equation}}
\newcommand{\bea}{\begin{eqnarray}}
\newcommand{\eea}{\end{eqnarray}}

\newcommand{\Tr}{\mbox{Tr}\;}
\newcommand{\tr}{\mbox{tr}\;}
\newcommand{\ket}[1]{\left|#1\right\rangle}
\newcommand{\bra}[1]{\left\langle#1\right|}

\newcommand{\avg}[1]{\left\langle #1\right\rangle}
\newcommand{\vnabla}{\mathbf{\nabla}}
\newcommand{\notes}[1]{\fbox{\parbox{\columnwidth}{#1}}}


\title{Generic framework for non-perturbative QCD in light hadrons}

\author{Wei-Yang Liu}
\email{wei-yang.liu@stonybrook.edu}
\affiliation{Center for Nuclear Theory, Department of Physics and Astronomy, Stony Brook University, Stony Brook, New York 11794-3800, USA}

\date{\today}

\keywords{QCD, gradient flow, instanton, topological charges, hadron, form factors}

\begin{abstract}

In this paper, we review several topological aspects of the QCD vacuum and recent progress on this quantitative framework for the low-lying hadron physics rooted in QCD by introducing the vacuum as a liquid of pseudoparticles. We have developed systematic density expansion on the dilute vacuum to calculate the vacuum expectation values (VEVs) and generalize the calculations to hadronic matrix element and hadronic form factors using the instanton liquid model (ILM). Thereby, the nonperturbative physics can be analyzed in a systematic framework with a few parameters: instanton size $\rho$ and instanton density $n_{I+A}$, and current quark mass $m$.


\end{abstract}

\maketitle


\section{Introduction}

Although hadron physics is firmly rooted in QCD, a theory over half a century old, the low-energy non-perturbative aspect is rather distinct from the high-energy realm, where the fundamental degrees of freedom in QCD are quarks and gluons. The task bridging the hadron physics to \textcolor{black}{quarks and gluons} has posed a significant challenge to physicists for many years. Among substantial progress in various directions, one key achievement is recognizing that nontrivial topological configurations in the vacuum play a pivotal role in understanding the non-perturbative aspects in hadrons from phenomenological studies \cite{ Diakonov:1985eg,Vainshtein:1981wh,Diakonov:1995ea,Miesch:2023hvl,Shuryak:2022wtk,Shuryak:2022thi,tHooft:1986ooh} (and references therein).

\begin{figure}
  \centering
    \includegraphics[width=\linewidth]{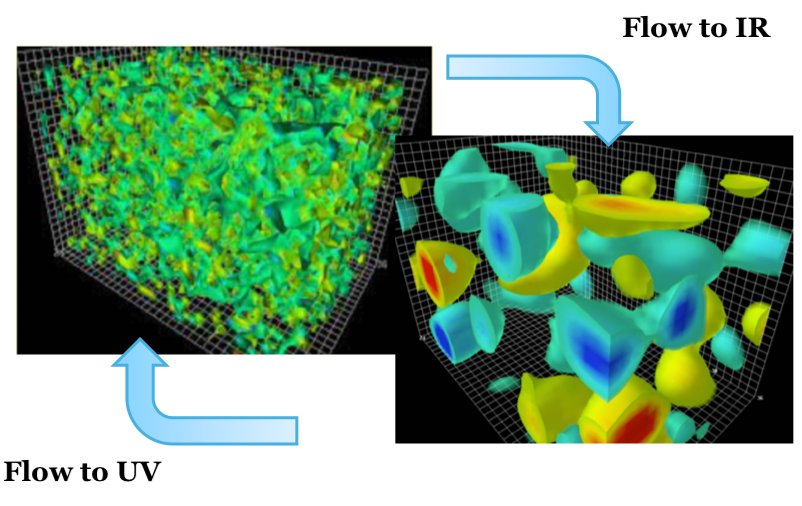}
    \caption{Visualization of the vacuum in gluodynamics, before cooling at a resolution of about $\frac 1{10}\,{\rm fm}$ (top),  and after cooling  at a resolution of about $ \frac 13\,{\rm fm}$ (bottom)~\cite{Moran:2008xq}, where the pseudoparticles emerge.}
    \label{fig:vacuum}
\end{figure}

From the lattice perspective, there is substantial evidence highlighting the importance of the topological structure \cite{Michael:1994uu,Michael:1995br,Leinweber:1999cw,Bakas:2010by,Biddle:2018bst,Hasenfratz:2019hpg,Athenodorou:2018jwu,Biddle:2020eec,Zimmermann:2024mar} and some even provide direct evidence for ILM \cite{Ringwald:1999ze,Faccioli:2003qz} (and references therein). 
Lattice QCD provides a first-principles framework for studying the non-perturbative regime of QCD. 
However, while it enables ab initio numerical calculations of hadronic observables, the numerical access alone does not always expose the underlying physical mechanisms responsible for these phenomena. 
In the continuum, such mechanisms are naturally tied to the structure of the non-perturbative QCD vacuum. 
The QCD instanton liquid model (ILM) 
offers \textcolor{black}{one of} the most compelling description of the underlying gauge configurations at low resolution, providing a physically transparent description of the relevant gauge-field configurations for understanding how vacuum topology shapes low-energy hadron structure. Therefore, it is crucial to revisit how the vacuum structure emerges within the lattice formulation. Gauge lattice configurations are heavily influenced by gluonic waves with wavelengths $\sim a$, the ultraviolet (UV) cutoff of the lattice. However, advanced renormalization techniques, such as the {\em gradient flow} procedure, effectively filter out these short-wavelength modes, uncovering the genuine non-perturbative fields that define the physical vacuum \cite{Moran:2008xq}. For a more comprehensive review of ILM, we refer the reader to \cite{Musakhanov:2023dsn, Schafer:1996wv,hutter2001instantons,Shuryak:2018fjr,Weiss:2025nzy,Diakonov:2002fq,Shuryak:1996wx}.

\subsection{Instanton liquid in gradient flow}

In Fig.~\ref{fig:vacuum}, after a few steps of cooling, the gluonic landscape resembles a rather dense ensemble of strongly correlated instanton-anti-instanton pairs. 
With continued cooling over longer flow times, these pairs gradually annihilate, leaving a more dilute ensemble of pseudoparticles that remain stable under further cooling.
For a comprehensive description of this procedure, we refer to the relevant \textcolor{black}{literature} \cite{Michael:1994uu,Michael:1995br,Leinweber:1999cw,Bakas:2010by,Biddle:2018bst,Hasenfratz:2019hpg,Athenodorou:2018jwu,Biddle:2020eec,Ringwald:1999ze} (and references therein). The detailed gradient flow (cooling) techniques have uncovered a remarkable semiclassical landscape composed of instantons and anti-instantons, the vacuum tunneling pseudoparticles with unit topological charges~\cite{Leinweber:1999cw}.

The key features of this landscape are~\cite{Shuryak:1981ff}
\begin{equation}
n_{I+A}\equiv \frac 1{R^4}\approx \frac 1{ {\rm fm}^{4}} \qquad\qquad\frac{ \rho}R \approx  \frac 13   \label{eqn_ILM}
\end{equation}
for the instanton plus anti-instanton density and size, respectively. The hadronic scale $R=1\,{\rm fm}$
emerges as the mean quantum tunneling rate of the pseudoparticles. 

\begin{figure}
  \centering
    \includegraphics[width=\linewidth]{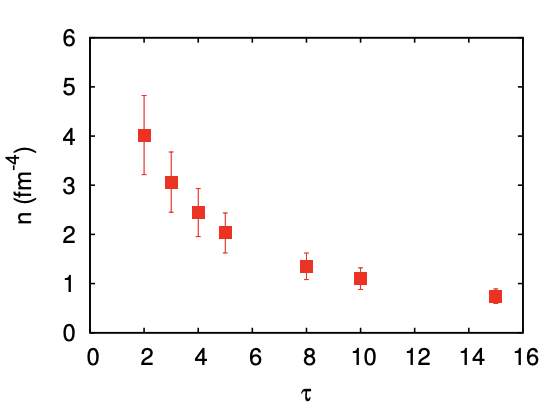}
    \caption{\textcolor{black}{Dependence of the density $n=n_{I+A}$ on the dimensionless gradient flow time $\tau=t/a^2$ with lattice spacing $a=0.046$ fm~\cite{Athenodorou:2018jwu}, determined by $\sqrt{8t_0}=0.1$ fm, or $1/\sqrt{8t_0}\sim2$ GeV. The quantum vacuum corresponds to the extrapolation $\tau\rightarrow0$}}
    \label{fig:COOL}
\end{figure}

In Fig.~\ref{fig:COOL}, we present the dependence of the 
instanton density $n$ on the cooling time $t$, as determined from the lattice analysis in \cite{Athenodorou:2018jwu}. The cooling time $t$ is related to the renormalization scale by
$$\mu\sim\frac{1}{\sqrt{8t}}$$ where the cooling time $t$ is defined in terms of the lattice spacing, $\tau=t/a^2$. Deep in the cooling time ($\tau=9$) or low resolution \textcolor{black}{$\mu=505~\mathrm{MeV}\sim 1/\rho$}, the tunnelings are sparse, well described by the ILM  with a dilute packing fraction 
\bea
\kappa_{I+A}\equiv \pi^2\rho^4 n_{I+A}\approx 0.1
\eea
where most instantons are annihilated with anti-instantons. This corresponds to the realm where the spontaneous breaking of chiral symmetry is commonly observed. At shorter cooling times ($\tau=0.6$) or high resolution $\mu=2~\mathrm{GeV}$, the larger density $n_{I+A}\sim 7.46/\rm fm^4$ is reached as more instanton-anti-instanton pairs are present.

The observed dramatic dropping of the instanton density in the gradient flow cooling can be primarily attributed to pair annihilation, leading to the equal decreasing rates of both $n_I$ and $n_A$. If we assume this is a first order process based on the collision picture, the flow time evolution of instanton and anti-instanton density will be given by

\begin{equation}
\frac{dn_I}{d\tau} = \frac{dn_A}{d\tau} = -\lambda(\tau) n_I(\tau) n_A(\tau)
\end{equation}
Here the rate constant $\lambda$ may vary with the flow time $\tau$ via the instanton size and inter-pseudoparticle distance. For simplicity, we assume that it
is well described by a constant. By assuming the initial condition $n_I=n_A=n_{I+A}/2$, we have \cite{Athenodorou:2018jwu}

\begin{equation}
n_{I,A}(\tau) = \frac{n_{I,A}(0)}{1 + \lambda n_{I,A}(0) \tau}
\end{equation}
where the numerical fitting of Fig.~\ref{fig:COOL} indicates $\lambda=0.1678$ fm$^4$.

\subsection{Topological configurations}

\begin{figure}
    \centering
    \includegraphics[height=4.2cm,width=\linewidth]{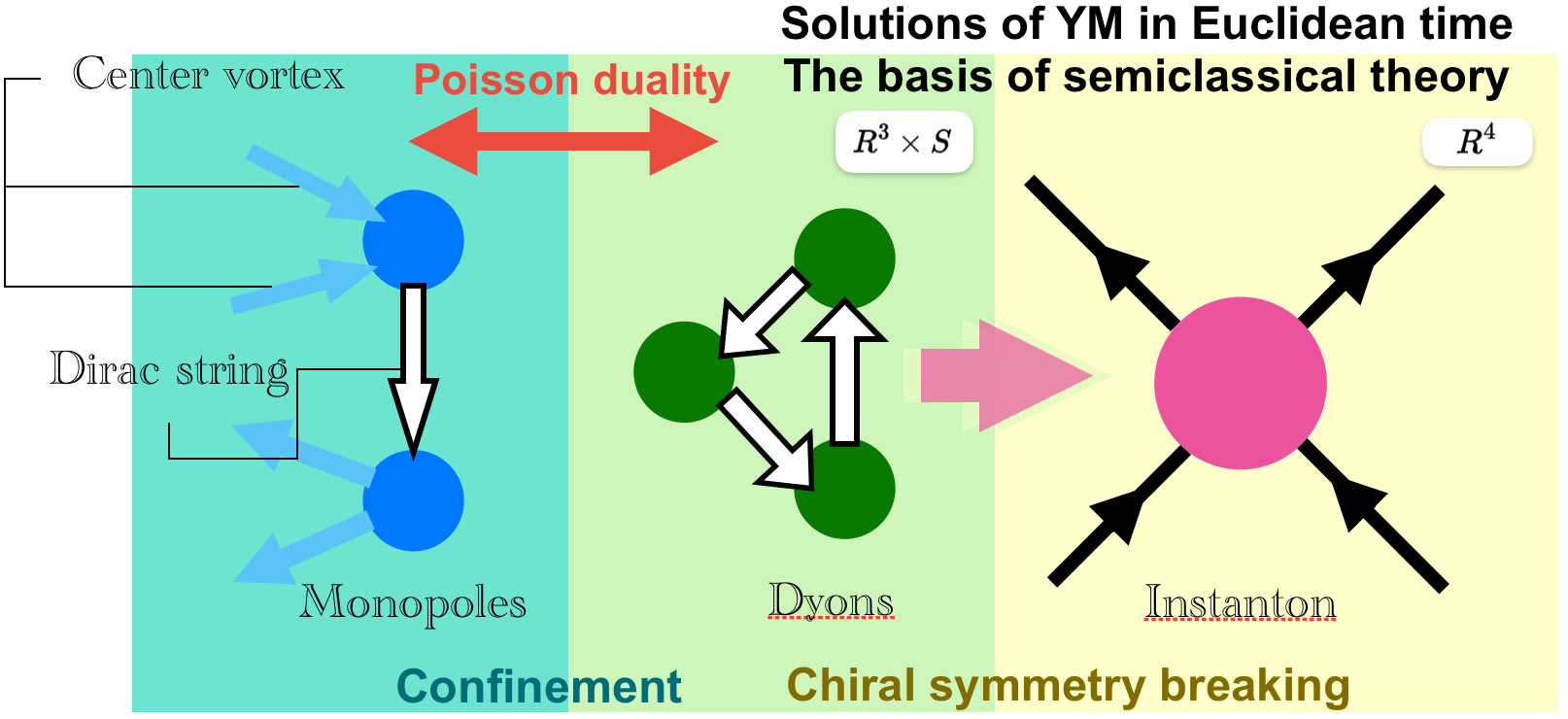}
    \caption{The relation between monopoles, dyons, and instantons. Monopoles are the endpoints of the center vortices, related to dyons (instanton-monopole) via Poisson transform at finite temperature. When the temperature cools down, dyons gets denser in the vacuum and eventually recombine to instantons at zero temperature where the full 4D vacuum degrees of freedom now is described by a dilute instanton liquid.}
    \label{fig:topo_field}
\end{figure}

The various topological gauge field configurations in the vacuum are deeply related to one another, as depicted in Fig.~\ref{fig:topo_field}. In a 4D gauge theory, monopoles are represented by world lines and center vortices are world sheets. When projected to a 3D subspace, monopoles are viewed as points and center vortices are lines. In $SU(2)$, the intersection of two center vortices where their fluxes merge, correspond to a monopole.  Each $SU(2)$ vortex carries a quantized flux of $\pi$. When two such vortices merge, their fluxes add to $2\pi$ (topologically trivial), which can be viewed as a thin flux tube terminating on a magnetic monopole. This string is referred to as a Dirac string.  Although it is not a physical observable due to the gauge dependence, it ensures the monopole consistently carries magnetic flux.
Although one may find effective model of monopoles and center vortices related to semi-classical description, they generally are not classical solution of Yang-Mills equation.

\subsubsection{Center P-vortices}

\begin{figure}
    \centering
    \includegraphics[width=1\linewidth]{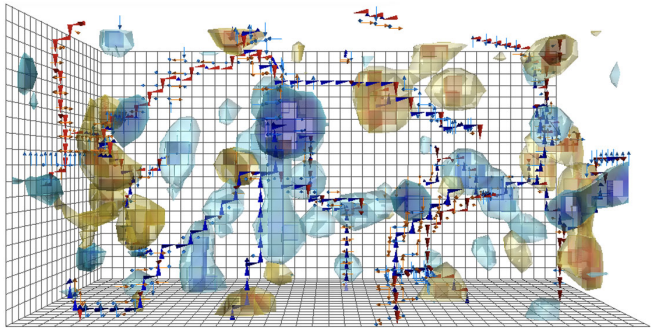}
    \caption{Instanton (yellow) and anti-instanton (blue) configurations in the deep-cooled Yang-Mills vacuum, threaded by center P-vortices using center projection on lattice \cite{Biddle:2019gke,Biddle:2020eec}. These topological configurations form the primordial gluon epoxy (hard glue) that underpins the origin of light hadron masses~\cite{Liu:2024rdm,Zahed:2021fxk} while the string-like center P-vortices play a key role in confinement, forming a world sheet in Euclidean time direction}
    \label{fig:p-v}
\end{figure}

The center P-vortex is a $Z_{N_c}$ flux tube configuration, characterized by a number of branching points (monopoles). The Wilson loop pierced by them picks up a non-trivial value from the group center $Z_{N_c}$. The analytical string-like structure of center vortices can be found in \cite{Diakonov:1999gg,Diakonov:2002bx,Maas:2005qt}. For instance, 

\begin{equation}
    \epsilon_{\mu\nu}n_\mu A_\nu^a(r) = \delta^{a3} \frac{\mu(r)}{r}
\end{equation}
where $n$ is the unit radial vector in the $xy$ plane with the radial coordinate $r=n\cdot x$ and $\mu(r)$ is the profile of the vortex with $\mu(0)=0$.  


In the case of $SU(3)$, as shown in Fig.~\ref{fig:p-v} with lattice simulations through gradient-flow smearing, the structure of the QCD vacuum is revealed as an intricate network of string-like center vortices, which behave as anchors of topological pseudoparticles. In Euclidean spacetime, these vortices form extended two-dimensional world sheets that fluctuate and percolate. Their essential feature is the presence of branching and recombination points, where multiple vortex sheets meet or split. In center-projected vacuum, these points act as sources and sinks of Abelian magnetic flux and appear as monopoles and anti-monopoles, reflecting genuine geometric singularities of the vortex surfaces.

 We further emphasize that this should be regarded as an interpretation of the lattice observations rather than as an established equivalence of the two descriptions

Strikingly, these branching regions are strongly correlated \textcolor{black}{with the locations of non-zero local
topological-charge density \cite{Biddle:2019gke,Biddle:2020eec}. The observed correlations suggest a close geometrical and topological connection between vortex structures and instanton-like topological configurations.} 
They both arise from localized regions where the gauge field configurations exhibit nontrivial topology due to rapidly varying configurations. In this picture, center-vortex sheets provide an underlying geometric structure that organizes topological charge. \textcolor{black}{When these sheets intersect, twist, or fold, they naturally generate localized topological lumps, which can be effectively identified as instantons \cite{Reinhardt:2002cm}.} 

\textcolor{black}{Although this should be regarded as an interpretation of the lattice observations rather than as a robust proof, this geometric interpretation unifies confinement and chiral symmetry breaking.} On one hand, confinement emerges from the percolation of extended vortex surfaces, which form a dense network throughout the vacuum. As these surfaces intersect at a Wilson loop, they induce fluctuating center phases that disorder its expectation value, leading to an area-law falloff characteristic of confinement. On the other hand, chiral symmetry breaking is driven by instantons, whose zero modes produce a near-zero Dirac spectrum. If monopole worldlines and instanton-like structures originate from the same vortex geometry, these phenomena are no longer independent but instead arise as complementary manifestations of a common underlying vacuum structure.

Many lattice results~\cite{Maas:2005qt,Langfeld:1998cz,Biddle:2019gke,Biddle:2020eec,Kamleh:2023gho} as well as theoretical studies~\cite{Nguyen:2024ikq,Nguyen:2025voy,Hayashi:2024yjc,Guvendik:2024umd,Greensite:2016pfc} have supported this assertion.

\subsubsection{Dyons}




\begin{figure}
    \centering
    \includegraphics[width=\linewidth]{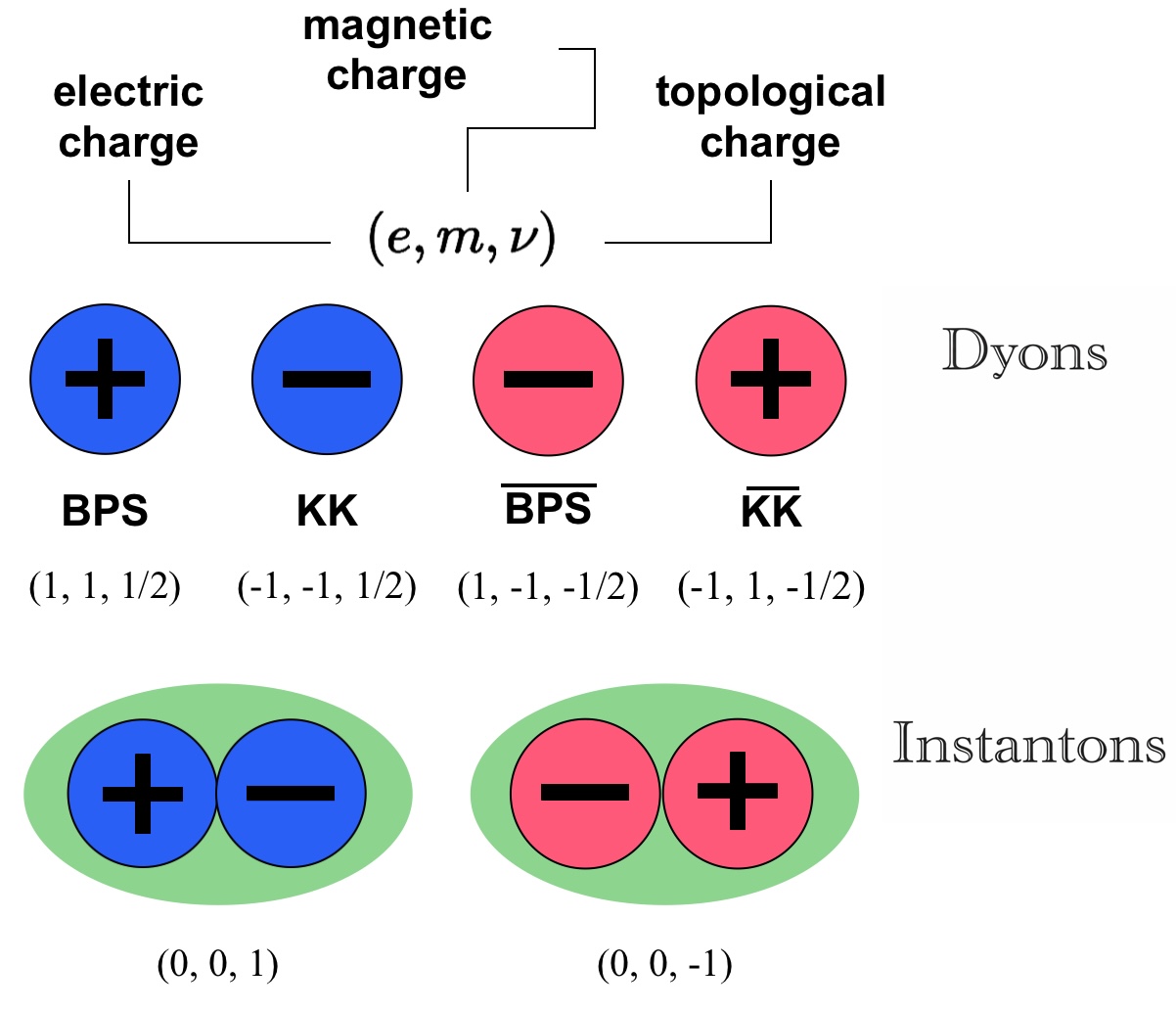}
    \caption{The constituents of instantons (calorons): BPS and KK dyons, and their anti-dyons with their electric, magnetc, and topological charges labeled}
    \label{charges}
\end{figure}
Instantons in Euclidean space $R^4$ can be generalized to finite temperature with a twisted temporal boundary condition defined on a circle $R^3\times S$. This modifies Belavin-Polyakov-Schwarz-Tyupkin (BPST) instanton solutions into Kraan-van Baal-Lee-Lu (KvBLL) instantons with non-trivial holonomy \cite{Kraan:1998pm,Lee:1998bb,Kraan:1998kp}. At finite temperature in $SU(N_c)$, each KvBLL instanton (caloron) with unit topological charge are decomposed into $N_c$ self-dual dyons \cite{Zhitnitsky:2006sr,Unsal:2008ch}. Each of them carries both electric, magnetic charges, and nonzero fractional topological charge specified by nontrivial Polyakov loop. This naturally extends the zero-temperature ensemble of instanton liquid to a finite-temperature dyon plasma ensemble characterized by long-range Coulomb-like interactions, which can be quantitatively described by the dyon liquid model (DLM) \cite{Liu:2015jsa,Liu:2015ufa,Diakonov:2009jq}.  DLM has been demonstrated to support a confining phase at sufficient dyon density, offering a comprehensive explanation for the confinement–deconfinement phase transition and chiral symmetry restoration. When the temperature cools down, dyons gets denser in the vacuum and eventually recombine to instantons at zero temperature where the full 4D vacuum degrees of freedom now is described by a dilute instanton liquid (ILM) (see Fig.~\ref{fig:topo_field}). These conclusions are supported numerically by \cite{Larsen:2015vaa,Larsen:2015tso} and by mean-field analyses \cite{Liu:2015jsa,Liu:2015ufa}. For a comprehensive overview, see \cite{Shuryak:2017kct}.

In $SU(2)$, the self-dual dyons with
electric and magnetic charges $(e, m)=(+, +)$ are called $M$, or Bogomolny–Prasad–Sommerfeld (BPS) dyons in \cite{Unsal:2007jx}. The ones with charges $(e, m)=(-, -)$ are called $L$, or Kaluza-Klein (KK) dyons. Their anti-self-duals $\bar M$ and $\bar L$ are the ones with $(e, m)=(+, -)$, and $(e, m)=(-, +)$, respectively. Their relation to instantons are illustrated in Fig.~\ref{charges}. In $SU(N_c)$, there are $N_c - 1$ BPS dyons (maximal
abelian subgroup), with charges counted from the Cartan generators, one KK dyon with the charge compensating BPS dyons to zero, and their anti-self-dual counterparts.

These dyon configurations are also closely related to monopoles, which arise as endpoints of center vortices.
They are related at finite temperature to the semiclassical description based on instanton–dyons through Poisson duality, even though monopoles themselves lack corresponding semiclassical theory description. In this dual perspective, dyon configurations can be viewed as quantum paths of moving and rotating monopoles in their collective coordinate space. Although the two descriptions are formally equivalent and encode the same underlying physics, instanton-dyon ensemble practically provide a more suitable description near and above the critical temperature $T\gtrsim T_c$ while monopole-based descriptions model the vacuum at much lower temperature $T\ll T_c$ \cite{Poppitz:2012sw,Dorey:2000qc,Poppitz:2011wy,Shuryak:2018fjr}, where $T_c$ is the critical temperature for QCD phase transition.

Consequently, at finite temperature, monopoles and instantons can be unified through a plasma picture of dyons (instanton-monopoles), which carry both topological and magnetic charges. In this picture, caloron are composed of multiple dyons, while monopoles are the dual description of them. This unified description becomes particularly relevant near the QCD phase transition, where dyon dynamics play a central role.

\subsection{Heavy quarkonia}

Although the heavy quark potential from instantons and anti-instantons rises linearly until about $1$ fm and flattens out at larger distances~\cite{Shuryak:2021fsu,Liu:2024sqj}, suggesting that those pseudoparticles are unlikely to account for the confinement, instanton effects may still contribute to heavy quarkonium systems despite the relative small size of heavy quarkonia. Those instanton effects are marginal but still essential for a thorough description of the heavy quarkonia spectra such as the static central potential and spin-dependent potential. These instanton-induced effects in heavy hadronic systems have been explored in~\cite{Musakhanov:2020hvk,Yakhshiev:2016keg,Shuryak:2021hng,Shuryak:2021fsu}. Throughout this paper, we will focus on the instanton effect for the low lying light hadrons only.

\subsection{Light hadrons}

In contrast to the heavy quarkonia, the property and dynamics of light hadrons should be tied to the vacuum structure. The major aspects of the QCD vacuum is the breaking of conformal symmetry and chiral symmetry \cite{Zahed:2021fxk,Zahed:2022wae,Diakonov:1995qy,Liu:2024rdm}, which govern the behavior of light hadrons at low energy. \textcolor{black}{The breaking of conformal symmetry is closely tied to the origin of most hadronic masses. 
In the instanton ensemble, this breaking is encoded in the form of weaker than Poisson fluctuations in the number of instantons and anti-instantons $N$ with the variance of $\sigma_t$, which defines the vacuum compressiblility. On the other hand, the breaking of $U(1)$ chiral symmetry is instead related to the topological charge $\Delta$ distributed in the form of Gaussian fluctuations with the variance of the topological susceptibility $\chi_t$, which is very sensitive to the presence of light quarks and vanishes in chiral limit \cite{Diakonov:1995qy,Schafer:1995pz}.}

The dynamical formation of quark condensates inside the QCD vacuum \cite{Diakonov:1995ea,Nowak:1989at} spontaneously breaks the chiral $SU(N_f)$ symmetry.

These pseudoparticles induce chiral symmetry breaking through fermionic zero modes with fixed chirality (left or right)~\cite{Diakonov:1995ea,Nowak:1989at}. As quarks pass through these pseudoparticles in the vacuum, they scatter, leading to the emergence of $2N_f$-fermi 't Hooft interaction. This interaction provides the QCD foundation for the Nambu–Jona-Lasinio (NJL) model, which effectively describes the dynamical formation of quark condensates and hadronic bound states. The bosonization of the 't Hooft interaction further results in the chiral Lagrangian at low momentum scales, unifying the low-energy light hadron dynamics of QCD under the framework of ILM.
 
\textcolor{black}{This paper reviews recent progress \cite{Liu:2024rdm,Liu:2024yqa,Liu:2024vkj,Liu:2024jno,Liu:2024sqj,Liu:2025kuc}} in extending ILM to calculate various hadronic form factors and partonic observables with a focus on light-front wave functions including quark and gluon content at low energy region where renormalization scale is expected to be $\mu\lesssim1$ GeV. This framework primarily employs in canonical ensemble and is extended to grand canonical frameworks when fluctuations in the instanton and anti-instanton ensemble play a pivotal role.

The organization of this paper is as follows. In Sec. \ref{QP}, we begin by reviewing the quark propagators \textcolor{black}{spanned in Dirac zero modes of} the multi-instanton vacuum structure with $1/N_c$ book-keeping planar resummation, showing the generation of dynamical constituent quark mass $M$ are tied to the quark condensate, leading to the spontaneous chiral symmetry breaking. 
In Sec.~\ref{ILM}, we review the development for a robust systematic theoretical framework which is based on Sec.~\ref{QP}. \textcolor{black}{In Sec.~\ref{Feyn}, we discuss the gluonic interactions in this instanton vacuum.} 
In Sec.~\ref{effective_Langrangian}, we proceed and derive the effective Lagrangian in the presence of a single instanton and an instanton–anti-instanton pair, and discuss their bosonization, which leads to chiral Lagrangian at low energy. In Sec.\ref{operators}, we establish the framework for calculating the VEVs of QCD operators as well as their hadron matrix elements, which is first introduced in \cite{Diakonov:1995qy,Weiss:2021kpt,Kacir:1996qn} and later extended to include higher order of instanton correlation in \cite{Liu:2024rdm}. All calculations of VEVs and hadron matrix elements are considered to be renormalized at the "intermediate scale" $\mu$, determined by the instanton size, with $\mu \sim 1/\rho \approx 0.6~\mathrm{GeV}$. 
It is important to distinguish this scale from the smaller "chiral" scales associated with the pion mass or the perturbative scales $\mu > 2$ GeV, where perturbative renormalization group (RG) evolution becomes applicable. In Sec.~\ref{grand}, we address the importance of the fluctuation in the instanton vacuum by extending the canonical ensemble of pseudoparticles to a grand canonical ensemble, to account for the fluctuations in their numbers which captures globally the scale and $U(1)$ anomalies. Finally, in Sec.~\ref{FF}, we summarize the full scale construction of the hadron form factors based on ILM perspective.

The appendices provide supplementary information as follows: Appendix \ref{App:conv} introduces the conventions used in this paper regarding the Euclidean QCD. Appendix \ref{App:singular} offers a concise review of the BPST instanton, including its definition and parametrization. In Appendix \ref{App:ZM}, the derivation of quark zero modes in a single instanton background is presented, along with several useful mathematical identities. \textcolor{black}{Appendix \ref{App:NZM} and \ref{App:reduction_large_time} elaborate the contributions of non-zero modes to the quark propagator and the corresponding LSZ reduction formula.} Finally, Appendix \ref{App:average} provides formula for color averaged integral of $SU(N_c)$ matrices with respect to the invariant Haar measure.

\section{Quark propagators in instanton vacuum}
\label{QP}

In this section, we introduce a systematic planar resummation to organize the quark propagation in various correlation functions in the multi-instanton background. 

\subsection{Quark Propagator}

The chiral symmetry breaking and the resulting dynamical generation of constituent quark masses in instanton vacuum explain various hadronic properties. More specifically, the pion, as a (pseudo) Goldstone boson, remains very light compared to constituent quark mass as a result of spontaneous chiral symmetry breaking. In contrast, the $ \rho $-meson exhibits a mass approximately twice of constituent quark mass, and the nucleon mass is about three times as large, indicating relatively weak binding.

For simplicity, we consider single flavor in the vacuum to begin with. The quark propagator $S(x,y)$ in the multi-instanton vacuum can be computed by the ensemble  average as \cite{Diakonov:1985eg,Pobylitsa1989TheQP}

\begin{widetext}
\begin{equation}
\begin{aligned}
\label{SRIV_1}
    S=&\left\langle\frac{1}{i\slashed{\partial}+\sum_I\slashed{A}_I+im}\right\rangle=\left\langle S_0+\sum_I(S_I-S_0)+\sum_{I\neq J}(S_I-S_0)S_0^{-1}(S_J-S_0)+\cdots\right\rangle
\end{aligned}
\end{equation}
\end{widetext}
where $S_I$ is the quark propagator with single instanton background defined as
\begin{equation}
    S_I=\frac{1}{i\slashed{\partial}+\slashed{A}_I+im}
\end{equation}
Here the ensemble average $\left\langle\cdots\right\rangle$ runs the entire instanton ensemble with sampling weighed by the interacting between the pseudoparticles (interaction instanton ensemble) or with the equal sampling (random instanton ensemble) for simplicity. 


\begin{equation}
    \langle \cdots \rangle=\prod_I\int \frac{d^4z_IdU_I}{V}\cdots
\end{equation}

\begin{figure}
    \centering
    \includegraphics[width=\linewidth]{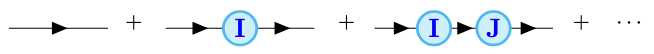}
    \caption{Quark propagator distorted by the instanton background with zero modes and non-zero modes. The zero mode contribution can be rewritten as 'tHooft vertices in \eqref{eq:ILM_vertices} due to the delocalization}
    \label{fig:Q_prop}
\end{figure}

This instanton expansion is presented graphically in Fig. \ref{fig:Q_prop}. Each instanton vertex $V_I$ denoted by blue circles can be obtained by Lehmann–Symanzik–Zimmermann (LSZ) reduction presented in \eqref{SI}, including both zero mode and non-zero mode contributions. 

\begin{equation}
\label{SI}
   V_I=S_0^{-1}(S_I-S_0)S_0^{-1} 
\end{equation}

The generalization to $N_f$ flavor vertex can be achieved straightforwardly by (See \cite{hutter2001instantons} for more details)

\begin{equation}
V_I=   \prod_{f}
   [S_0^{-1}(S^f_I-S_0)S_0^{-1}]\ =\
   \begin{tikzpicture}[scale=0.48,baseline=(z)]
   \begin{feynhand}
   \path (0,0) -- (4,0);
    \vertex (a) at (0,0){};
    \vertex [particle] (b) at (-2,1) ;
    \vertex [particle] (c) at (-2,0) ;
    \vertex [particle] (e) at (-2,-1) ;
    \vertex [particle] (d) at (2,1) ;
    \vertex [particle] (f) at (2,0) ;
    \vertex [particle] (g) at (2,-1) ;
    \vertex [particle] (z) at (1,0);
    \propag [fer] (c) to (a);
    \propag [fer] (a) to (f);
    \propag [fer] (b) to (a);
    \propag [fer] (a) to (d);
    \propag [fer] (e) to (a);
    \propag [fer] (a) to (g);
    \filldraw[fill=cyan!1, color=cyan!20, draw=cyan, very thick] (0,0) circle (0.5);
    \node at (0, 0) {\textcolor{blue}{\bf I}};
   \end{feynhand}
   \end{tikzpicture}
\end{equation}
This vertex becomes 't Hooft vertex in \eqref{eq:tHooft} when only zero mode is considered. For simplicity, here our discussion will stick to the single flavor case. Large $N_c$ QCD is essentially a quenched approximation dominated by planar graphs. The same applies to its semi-classical version in terms of a random instanton vacuum. As a result, in the large $N_c$ limit, the resummation of these planar diagrams repackages the diagrams involving the same instanton at both the beginning and the end, yielding ~\cite{Pobylitsa1989TheQP,Liu:2023yuj}


\begin{equation}
\begin{aligned}
S=&\left\langle S_0+S_0\left(\sum_IM_I\right)S_0+\cdots\right\rangle\\
=&\left\langle\frac1{S_0^{-1}-\sum_IM_I}\right\rangle
\end{aligned}
\end{equation}
where the effective quark self-energy in instanton vacuum is given by the iterative equation
\begin{equation}
\begin{aligned}
\label{MqI}
    M_I
    =&V_I+V_I(S-S_0)M_I
\end{aligned}
\end{equation}


By defining the averaged total quark self-energy, 
\begin{equation}
    i\sigma=-\sum_I\int \frac{d^4z_IdU_I}{V}M_I
\end{equation}
the solution to \eqref{MqI} yields a gap-like equation

\begin{equation}
\begin{aligned}
   &i\sigma=\frac{n_{I+A}}{N_c}\int d^4z_I\mathrm{Tr}_c\left(\slashed{A}_I\frac{1}{S_I^{-1}+i\sigma}\left(S_0^{-1}+i\sigma\right)\right)
\end{aligned}
\end{equation}
which can be solved order by order with the density expansion $\sigma=\sqrt{\frac{n_{I+A}}{N_c}}\sigma_0+\frac{n_{I+A}}{N_c}\sigma_1+\cdots$. The dependence on square root of density $\sqrt{n_{I+A}/N_c}$ at the leading order is the consequence of disordering in the random instanton vacuum. 

Eventually, a momentum-dependent constituent quark mass naturally emerge in instanton vacuum, determined by  \cite{Liu:2023fpj,Kock:2020frx,Pobylitsa1989TheQP}

\begin{eqnarray}
    M(k)=m+\sigma(k)
\end{eqnarray}

The nearly massless quarks acquire a substantial dynamical mass, denoted as $ M(k) $. The quark propagator can be written as
\begin{equation}
\label{running mass}
    S(x,y)=\int \frac{d^4k}{(2\pi)^4}\frac{\slashed{k}-iM(k)}{k^2+M^2(k)}e^{-ik\cdot (x-y)}
\end{equation}

\subsection{Delocalization in zero modes}

  
Generally, quark propagator with single instanton $S_I$ appears as a sum over zero modes and non-zero modes. Yet in the case of light quarks, the zero modes dominates due to the nearly zero (current) mass. The propagator in the single instanton can be approximated by \cite{Diakonov:1985eg}

\begin{equation}
    S_I(x,y)\simeq\frac{\phi_I(x)\phi_I^\dagger(y)}{im}+S_0(x-y)
\end{equation} 

The non-zero mode contribution is smeared into a free propagator $S_0$. In this smearing treatment, the propagator appears in the instanton resummation \eqref{SRIV_1} can be simplified~\cite{Diakonov:1985eg,Kock:2020frx,Kock:2021spt}

\begin{equation}
\begin{aligned}
\label{eq:Q_prop}
    S(x,y)&\simeq S_0(x-y)\\
    &+\left\langle\sum_{I,J}\phi_I(x)\frac{1}{im-im D_{IJ}-T_{IJ}}\phi_J^\dagger(y)\right\rangle
\end{aligned}
\end{equation}

The hopping between the \textcolor{black}{pseudoparticles} with the opposite and the same duality are defined by $T_{IJ}$ and $D_{IJ}$ respectively with their conjugation representing the reverse process.
\begin{equation}
\begin{aligned}
\label{TIJ}
    T_{IJ}(u,R)=&\int d^4x\phi_I^\dagger(x) i\slashed{\partial}\phi_J(x)\\
=&-4\pi^2\rho^2\mathrm{Tr}_c(\tau_\mu^+u)\frac{R_\mu}{R}\frac{dT(R)}{dR}
\end{aligned}
\end{equation}

\begin{equation}
\begin{aligned}
\label{DIJ}
    D_{IJ}(u,R)=&\int d^4x\phi_I^\dagger(x)\phi_J(x)-\delta_{IJ}\\
    =&4\pi^2\rho^2\mathrm{Tr}_c(\mathds{1}_2u)T(R) -\delta_{IJ}
\end{aligned}
\end{equation}
where $\mathds{1}_2$ is the $N_c\times N_c$ diagonal matrix defined by $\mathrm{diag}(1,1,0,\cdots,0)$ and the hopping integral $T(R)$ is defined
\begin{equation}
\begin{aligned}
\label{eq:hop}
    T(R)=&\int\frac{d^4k}{(2\pi)^4}\frac{\mathcal{F}(\rho k)}{k^2}e^{-ik\cdot R}\\
    =&\frac{1}{2\pi^2R}\int_0^\infty dk\mathcal{F}(\rho k)J_1(k R)
\end{aligned}
\end{equation}
with the quark zero mode non-local form factor $\mathcal{F}(k)$ defined in \eqref{ZMform} and $J_n$ Bessel functions of the first kind. The relative distance is typically $R=z_I-z_J\sim\rho<2\rho$. The hopping between the pseudoparticles with the same duality are strongly suppressed by the quark current mass $m$.

If we only focus on the contribution from zero modes, the iterative equation in \eqref{MqI} can be solved by ansatz assuming quark self-energy $M_I$ is of the form,

\begin{equation}
\label{self_energy_q}
M_I=\frac{1}{i\sigma_0}S_0^{-1}|\phi_I\rangle\langle\phi_I|S_0^{-1}
\end{equation}
The solution to \eqref{MqI} can be simplified to a self-consistent condition for the disordering mass parameter $\sigma_0$.

\begin{equation}
\begin{aligned}
\label{eq:det_gap}
    &\sigma_0
    =m\\
    &+8\pi^2\rho^2\int\frac{d^4k}{(2\pi)^4}\frac{\sigma(k)\left[1+\frac{m}{k^2}(m+\sigma(k))\right]}{k^2+[m+\sigma(k)]^2}\mathcal{F}(\rho k)
\end{aligned}
\end{equation}

A comparison between $M_I$ and $V_I$ reveals that the zero-mode singularity $1/m$ is regularized by the inclusion of disordering effects, leading to a finite contribution $1/\sigma_0$ even in chiral limit. By substituting the disordering $\sigma_0$ in \eqref{eq:det_gap} back into \eqref{self_energy_q} 
the constituent mass is obtained as

\begin{equation}
\label{eq:cons_m}
    M(k)\simeq m+\frac {n_{I+A}}{2N_c} \frac{4\pi^2\rho^2}{\sigma_0}\mathcal{F}(\rho k)
\end{equation} 
where $\mathcal{F}(\rho k)$ is the profile of the quark zero mode~\cite{Liu:2023yuj,Kock:2020frx} defined by
\begin{equation}
\label{ZMform}
    \sqrt{\mathcal{F}(k)}=-z\frac{d}{dz}[I_0(z)K_0(z)-I_1(z)K_1(z)]\bigg|_{z=\frac{\rho k}{2}}
\end{equation}

\begin{figure}
    \centering
    \includegraphics[width=1\linewidth]{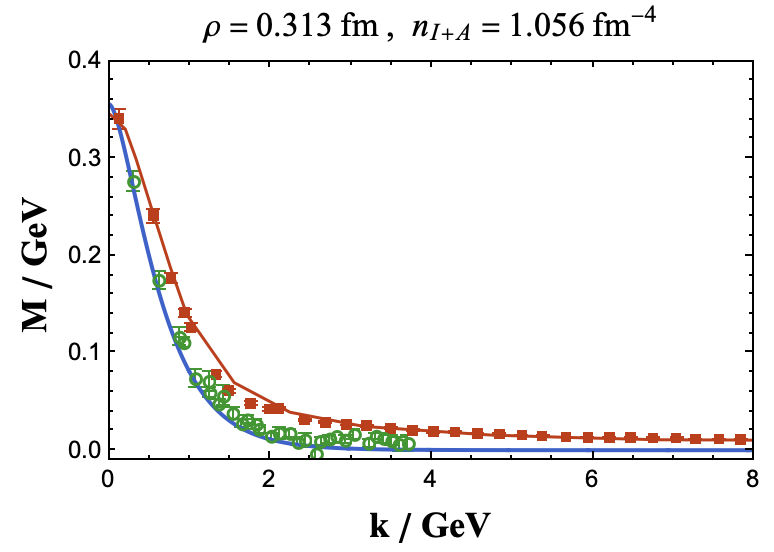}
    \caption{The constituent mass $M(k)$ running with the quark momentum $k$ with the instanton size $\rho=0.313$ fm and $n_{I+A}=1.056$ fm$^{-4}$ compared with lattice QCD using dynamical $O(a)$-improved Wilson fermions \cite{Oliveira:2018lln} (red) and result using overlap and Asqtad fermions \cite{Bowman:2004xi} in Landau gauge (green).}
    \label{fig:constituent_mass}
\end{figure}


The quark condensate in chiral limit at the leading order of instanton density reads
\begin{equation}
\begin{aligned}
\label{qq}
    \langle \bar q q\rangle=&-4N_c\int\frac{d^4k}{(2\pi)^4}\frac{M(k)}{k^2+M^2(k)}\\
    \simeq&-\frac{n_{I+A}}{\sigma_0}+\mathcal{O}(n_{I+A}^2)
\end{aligned}
\end{equation}

In Fig.~\ref{fig:constituent_mass}, we compare our result of constituent mass with the lattice QCD approach \cite{Oliveira:2018lln,Bowman:2004xi}.

This planar resummation in the multi-instanton vacuum with $1/N_c$ book-keeping can be straightforwardly generalized to any correlation functions. For more details, see \cite{Pobylitsa1989TheQP,hutter2001instantons}. 

\subsection{Non-zero mode dominance in heavy quarks}

Since the zero modes depend inversely on the quark mass, they mostly do not contribute to heavy flavor. Instead, heavy quarks receive significant contribution from non-zero modes~\cite{Chernyshev:1995gj}. The heavy quark propagator moving in velocity $v_\mu$ can be written in the form of Wilson line as,

\begin{equation}
\begin{aligned}
S(x)=\frac{1+\slashed{v}}2\delta^3(\vec{x})\Theta(\tau)\left\langle\mathcal{P}\exp\left(i\int d\tau v_\mu A_\mu\right)\right\rangle
\end{aligned}
\end{equation}
where $\vec{x}$ denotes the transverse space perpendicular to $v_\mu$.

In this heavy mass limit, the multi-instanton contribution in propagator resums to Wilson line where \textcolor{black}{the} full instanton contribution to it is given by the exponent of the all-order single instanton result~\cite{Shuryak:2000df,Shuryak:2021hng,Dorokhov:2002qf}

\begin{equation}
\begin{aligned}
\label{loop_inst}
    &\left\langle\mathcal{P}\exp\left(i\int dx_\mu A_\mu\right)\right\rangle\\
    &=\exp\left[\frac{1}{N_cV}\sum_I\int d^4z ~\mathrm{Tr}_c\left(W_I(\rho,z)-1\right)\right]
\end{aligned}
\end{equation}
with the single instanton inserted Wilson line
\begin{equation}
\begin{aligned}
&W_I(\rho,z)=\\
&\exp\left(i\tau^a\int_{C} dx_\mu  \frac{\bar\eta^a_{\mu\nu}(x-z)_\nu\rho^2}{(x-z)^2[(x-z)^2+\rho^2]}\right)
\end{aligned}
\end{equation}
Therefore, for the heavy quark propagator, the non-zero mode contribution thus can be studied by the straight line along $v_\mu$. The result reads

\begin{equation}
    \left\langle\mathcal{P}\exp\left(i\int_{-\infty}^{\infty} d\tau v\cdot A_I\right)\right\rangle=\cos \phi +i\frac{\vec{x}}{|\vec{x}|}\cdot\vec{\tau}\sin\phi
\end{equation}
where instanton cummulated phase is 
\begin{equation}
    \phi=\pi\left(1-\frac{|\vec{x}|}{\sqrt{|\vec{x}|^2+\rho^2}}\right)
\end{equation}

\section{Theory of instanton liquid ensemble}
\label{ILM}
For a more quantitative description of the light quarks in QCD vacuum at low resolution, 
we will focus on the pseudoparticles illustrated in Fig.~\ref{fig:vacuum}, designated by $N_+$ the number of pseudoparticles and $N_-$ the number of pseudoparticles with opposite charges.  For fixed numbers $N_\pm$, the canonical partition function $Z_{N_\pm}$ is
\begin{equation}
\begin{aligned}
\label{eq:Z_N}
   Z_{N_\pm}=&\frac{1}{N_+!N_-!}\int\prod_{I=1}^{N_++N_-}d\Omega_I n_0(\rho_I)\rho_I^{N_f}e^{-S_{int}}\\
   &\times
  \prod_{f}\mathrm{Det}(\slashed{D}+m_f)_{\mathrm{low}}
\end{aligned}
\end{equation}
where $d\Omega_I=d\rho_I d^4z_I dU_I$ is the conformal measure (size $\rho_I$, center $z_I$, and color orientation $U_I$) for each single (anti-)instanton and $S_{int}$ is the gauge interaction among pseudoparticles. The mean tunneling rate (one-loop) is
\begin{equation}
\label{eq:n}
n_0(\rho)=C_{N_c}(1/\rho^5)\left(8\pi^2/g^2\right)^{2N_c} e^{-8\pi^2/g^2(\rho)}
\end{equation}
with $C_{N_c}$ is the number dependent on color number $N_c$ defined as 
$$
C_{N_c} = \frac{0.466 \exp(-1.679 N_c)}{(N_c - 1)! (N_c - 2)!}
$$

At two-loop, the renormalization group requires the inverse coupling $8\pi^2/g^2(\rho)$ in the exponent to run with the Gell-Mann-Low beta function. The size distribution (two-loop) is \cite{Schafer:1996wv}
\begin{equation}
\begin{aligned}
\label{eq:dn2}
&n_0(\rho)=\frac{C_{N_c}}{\rho^5}[S_1(\rho)]^{2N_c}\\
&\times\exp\left[-S_2(\rho)
+ \left(2 N_c - \frac{b_1}{2 b_0}\right)
\left(\frac{b_1}{2 b_0}\right)
\frac{\ln S_1(\rho)}{S_1(\rho)}\right]
\end{aligned}
\end{equation}
where the one-loop running inverse coupling is given by $S_1=-b_0\ln(\rho\Lambda)$ and two-loop $S_2$ is given by
\begin{equation}
\label{eq:run_coup}
S_2(\rho)=S_1(\rho)
+ \frac{b_1}{2\,b_0}
\ln\!\left(
\frac{2}{b_0}\,S_1(\rho)
\right)
\end{equation}
with the beta function coefficient $b_0=(11 N_c - 2 N_f)/3$ and $b_1=(34 N_c^2 - 13 N_c N_f + 3 N_f/N_c)/3$
that appear in the Gell-Mann-Low beta function defined by
\begin{equation}
\label{eq:beta_GL}
\beta(g^2)=\mu\frac{\partial g^2}{\partial\mu}=-\frac{b_0g^4}{8\pi^2}-\frac{b_1g^6}{2(8\pi^2)^2}+{\cal O}(g^8).   
\end{equation}
Here $N_f=0$ is set for pure gluon vacuum. 

\textcolor{black}{In general the inverse coupling in the exponential and the prefactor runs at different loop order. This is because the exponential and prefactor in the instanton density arise from different pieces of the semiclassical expansion. The exponential $e^{-8\pi^2/g^2(\rho)}$ is determined by the classical instanton action and is naturally RG-improved using the running coupling, whose scale dependence is governed by the RG and can be included to higher-loop accuracy. In contrast, the prefactor originates from the functional determinant of quantum fluctuations around the instanton background, that is, Gaussian integration over nonzero modes and zero-mode normalization. This part is only known to one-loop order~\cite{tHooft:1976snw}, and higher-order corrections remain to be computed. As a result, the two terms are generally improved by RG separately.}


The fermion determinant receives contribution from the high momentum modes
as well as the low momentum modes. The contribution of the higher modes are localized on the pseudoparticles. They normalize the mean-density rate,  with an additional factor of $\rho^{N_f}$. The low momentum modes in the form of quasi-zero modes, are delocalized among the pseudoparticles. Therefore, in ILM, the fermionic determinant is usually represented by the determinant of the overlap matrix $T_{IA}$ in the zero mode subspace, which can be rewritten by effective vertices $\Theta_{I}$ \cite{Diakonov:1995ea, Diakonov:1995qy,Schafer:1996wv,doi:10.1142/1681}. Now, the generic 't Hooft vertices read 

\begin{equation}
    \begin{aligned}
        \label{eq:tHooft}
        &\Theta_{I}=\prod_f\left[m_f-\frac{4\pi^2\rho^2}{8}\bar{\psi}_fU_I\tau^\mp_\mu\tau^\pm_\nu\gamma_\mu\gamma_\nu U_I^\dagger\frac{1\mp\gamma^5}{2}\psi_f\right]
    \end{aligned}
\end{equation}
to lowest order in the current quark masses $m_f$. The emergent vertices (\ref{eq:tHooft}) can be generalized to include further finite size effects of the pseudoparticles. More specifically, each  quark field in the interaction vertices $\Theta_{I}$ get dressed
\begin{align}
&\psi(k)\rightarrow\sqrt{\mathcal{F}(\rho k)}~\psi(k)
\end{align}
with non-local quark form factor
which is essentially the profiling of the instanton by the quark zero mode. 

With the instanton numbers fixed, the QCD path integral can be rewritten as

\begin{widetext}
\begin{equation}
\label{eq:Z_N2}
   Z_{N_\pm}=\int \mathcal{D}\psi \mathcal{D}\bar\psi \frac1{N_+!N_-!}\prod_{I=1}^{N_++N_-}\left(\int d^4z_IdU_I\int d\rho_I n_0(\rho_I)\rho_I^{N_f}\Theta_I(z_I) \right)e^{-S_{int}}\exp\left(-\int d^4x\bar{\psi}\slashed{\partial}\psi\right)\\
\end{equation}
\end{widetext}

\subsection{Instanton size distribution}
\label{Sec:inst_size}
The instanton size distribution of the pseudoparticles is well captured semi-empirically by the original ILM \cite{Shuryak:1981ff},  confirmed then by various mean-field studies  \cite{Diakonov:1995ea,Nowak:1996aj}. The small size distribution follows from the conformal nature of the instanton moduli and perturbation theory. The large size distribution is non-perturbative, but cut-off by $R$, the mean separation of the instantons (anti-instantons) in the vacuum.  Thus, the size distribution has been proposed in a specific form that reads \cite{Schafer:1996wv}

\begin{equation}
\label{dn_dist}
n_\pm(\rho) =n_0(\rho)\, e^{-2\pi\sigma\rho^2}
\end{equation}
with $n_0$ is the quenched instanton density defined in \eqref{eq:n} (one-loop) and \eqref{eq:dn2} (two-loop).

\begin{figure}
    \centering
\subfloat[\label{fig:size_distSU2}]{\includegraphics[width=1\linewidth]{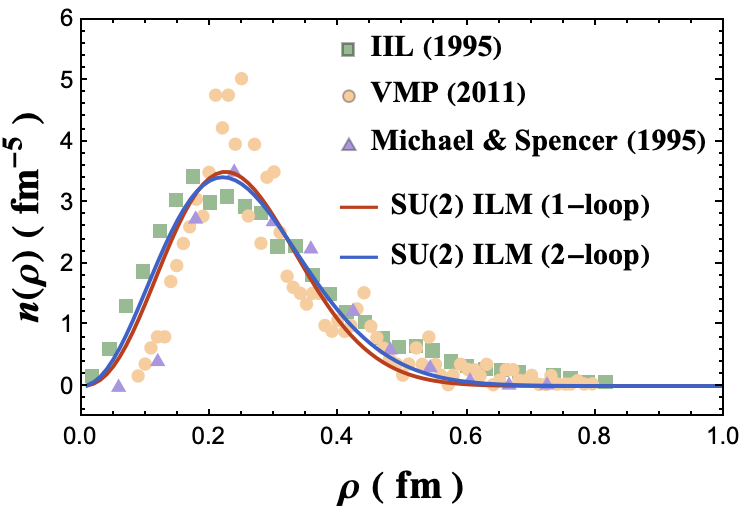}}
 \hfill
\subfloat[\label{fig:size_distSU3}]{\includegraphics[width=1\linewidth]{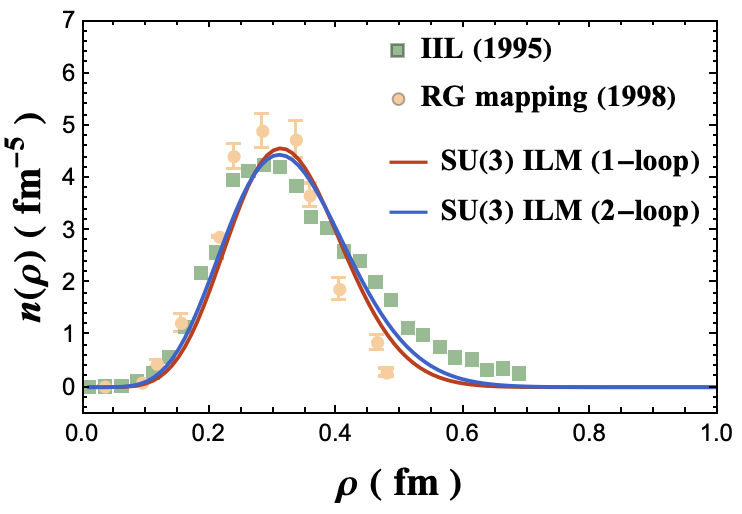}}
    \caption{\textcolor{black}{(a) $SU(2)$ instanton size distribution compared with IIL and lattice results, normalized to $n_{I+A}=0.93~\mathrm{fm}^{-4}$. See text.
(b) $SU(3)$ instanton size distribution compared with IIL, RG mapping, and UKQCD lattice results, normalized to $n_{I+A}=1.07~\mathrm{fm}^{-4}$. See text.}}
    \label{fig:size_dist}
\end{figure}

In the present treatment, only the small size distribution follows from the conformal nature of the instanton moduli and perturbation theory, while the large size distribution is intrinsically nonperturbative. A cut-off scaled by $R$, the mean separation of the instantons (anti-instantons) in the vacuum must be introduced. Otherwise, the integration over the size distribution leads to IR divergences. Phenomenologically, this suppression can be attributed to a dual monopole condensate, as inferred from the flux-tube tension $\sigma=(0.44~\mathrm{GeV})^2\sim1/R^2$~\cite{Shuryak:1999fe,Dorokhov:2002qf}.
Similarly, the statistical simulations of the ensemble \cite{Shuryak:1995pv} suggest a quadratic $\rho$ dependence as well. 

Table~\ref{tab:inst_dist_para} estimates the instanton mean size $\rho$ and the density $n_{I+A}$ using \eqref{dn_dist} fitted with IIL ensemble \cite{Shuryak:1995pv}. The result is well consistent with the estimation on lattice \cite{Millo:2011zn,Hasenfratz:1998qk} for both $N_c=2$ and $N_c=3$ while the result calculated by UKQCD group shows larger mean size $\rho=0.5$ fm \cite{Smith:1998wt}.

\begin{table}
    \centering
    \begin{tabular}{c|c|c|c}
    \hline
    & $\rho$ & $n_{I+A}$ & $\sigma$ \\
    \hline
        $SU(2)$ & $0.254(5)$ fm & $0.93(3)$ fm$^{-4}$ & $(0.271(7)\ \mathrm{GeV})^2$  \\
        $SU(3)$ & $0.328(4)$ fm & $1.07(3)$ fm$^{-4}$ & $(0.323(5)\ \mathrm{GeV})^2$ \\
    \hline
    \end{tabular}
    \caption{The instanton mean size and density obtained by fitting the results in IIL ensemble \cite{Shuryak:1995pv} with \eqref{dn_dist}.  \textcolor{black}{The classical string tension $\sigma$ is estimated with $\Lambda=0.3053(24)$ GeV for $SU(2)$ and $\Lambda=0.3246(25)$ GeV for $SU(3)$.}}
    \label{tab:inst_dist_para}
\end{table}

\textcolor{black}{Our ILM results on the instanton size distribution are presented in Fig.~\ref{fig:size_dist}. In Fig.~\ref{fig:size_distSU2}, $SU(2)$ instanton size distributions ($N_f=0$) in 1-loop (solid red) and 2-loop (solid blue) are fitted with the result from IIL ensemble (green square) \cite{Shuryak:1995pv} using~\eqref{dn_dist} and compared to lattice calculation using vacuum manifold projection (VMP) with lattice spacing $a=0.17$ fm on a $16^4$ lattice (green dot) \cite{Millo:2011zn} and under-relaxed cooling algorithm with combination of $a=0.12$ fm on a $16^4$ lattice 
and $a=0.08$ fm on a $24^4$ lattice (purple triangle) \cite{Michael:1995br}. All distributions presented in Fig.~\ref{fig:size_distSU2} is normalized to the instanton density $n_{I+A}=0.93$ fm$^{-4}$ which is obtained via fitting. In Fig.~\ref{fig:size_distSU3}, $SU(3)$ instanton size distributions ($N_f=0$) in 1-loop (solid red) and 2-loop (solid blue) are fitted with the result from IIL ensemble (green square) \cite{Shuryak:1995pv} using Eq.~\eqref{dn_dist}, compared to RG mapping method on lattice (orange circle) \cite{Hasenfratz:1998qk}. All distributions presented in Fig.~\ref{fig:size_distSU3} is \textcolor{black}{normalized to} instanton density $n_{I+A}=1.07$ fm$^{-4}$ which is obtained via fitting.}

The ILM predictions agree with lattice calculations using VMP \cite{Millo:2011zn} ($N_c=2$) and RG mapping \cite{Hasenfratz:1998qk} ($N_c=3$). The results indicate that small instantons are more suppressed in $SU(3)$ than $SU(2)$, consistent with the ILM prediction $n_\pm(\rho) \sim \rho^{\frac{11}{3}N_c - 5}$. On the other hand, in \cite{Smith:1998wt}, the density at large distances was found to decrease as $1/\rho^{11}$, while for small instantons, it scales as $\rho^6$, which agrees with our ILM prediction on small instanton density $n_\pm(\rho) \sim \rho^{\frac{11}{3}N_c - 5}$ for $N_c = 3$.  

Both phenomenological evidence and available lattice data suggest that instantons larger than $\rho \simeq 1/3$ fm are significantly suppressed in QCD. This observation cannot be explained by the leading-order semi-classical formula. This suppression can be attributed to essentially three possibilities: the instanton distribution may be regulated by higher-order quantum effects, by classical instanton interactions, or by the interaction of instantons with other classical objects (e.g., monopoles or strings) \cite{Schafer:1996wv}.

\subsection{Emergent $^\prime$t Hooft vertices}
\label{sec:tH}
\begin{figure}
\centering
\subfloat[\label{fig:tHooft_3}]{\includegraphics[width=0.33\linewidth]{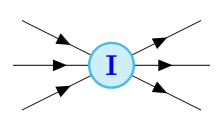}}
\hfill
\subfloat[\label{fig:tHooft_2}]{\includegraphics[width=0.33\linewidth]{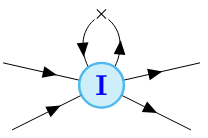}}
\hfill
\subfloat[\label{fig:tHooft_1}]{\includegraphics[width=0.33\linewidth]{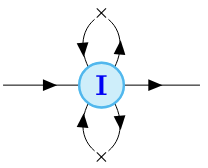}}
\caption{Feynmann diagrams in the ILM. (a) $2N_f$-quark 't Hooft vertices induced by zero modes. (b) one-flavor reduction by mean field mass insertion $m^*/(4\pi^2\rho^2)$. (c) two-flavor reduction.}
\label{fig:ILM}
\end{figure}

Each emerging vertex $\Theta_{I}$ in (\ref{eq:Z_N2}) is randomly averaged over the single pseudoparticle moduli with mean size fixed,

\begin{equation}
\label{eq:ILM_vertices}
  \theta_\pm(x)=\int dU_{I}\Theta_{I}
\end{equation}
The resulting effective instanton vertices are composed of the $2N_f$-quark 't-Hooft interaction ('t-Hooft Lagrangian).  Thus, in ILM, the quark degrees of freedom and their dynamics are mostly encoded in these vertices $\theta_\pm$. The typical 't Hooft vertices involving three flavors are shown in Fig.~\ref{fig:tHooft_3}.
Only different flavor can pass through the same instanton simultaneously due to Pauli exclusion. In Fig.~\ref{fig:tHooft_2} and \ref{fig:tHooft_1}, the flavor reduction can be achieved by looping up some of the flavors with the insertion of effective quark mass $m^*$ (See Sec.~\ref{sec:mdet}). 

In the thermodynamic limit ($V\rightarrow\infty$ with $n_{I+A}$ fixed) along with the large $N_c$ limit (the size of instanton is fixed by the small mean value $n_\pm(\rho)\rightarrow \delta(\rho-\bar\rho) n_{I+A}/2$), the emergent vertices $\Theta_I$ is exponentiated around the saddle point of the partition function $Z_{N_\pm}$ in \eqref{eq:Z_N2}, giving
\begin{equation}
\label{ZNEFF}
   Z_{N_\pm}\propto
   \int \mathcal{D}\psi \mathcal{D}\psi^\dagger \exp\left(-\int d^4x\mathcal{L}_{\mathrm{eff}}\right)
\end{equation}
where the effective Lagrangian in Euclidean space reads \cite{Liu:2024rdm,Diakonov:1995qy}

\begin{equation}
\begin{aligned}
\label{eq:effective_action}
    \mathcal{L}_{\mathrm{eff}}=-\psi^\dagger i\slashed{\partial}\psi &-\frac{N}{2V}
    \left(\frac{1}{m^*}\right)^{N_f}(1+\delta)\theta_+\\
    &-\frac{N}{2V}
    \left(\frac{1}{m^*}\right)^{N_f}(1-\delta)\theta_-
\end{aligned}
\end{equation}

 The explicit form of the Lagrangian can be found in Sec \ref{tHooft_L}. The emergent parameters $G_I$ and $\delta$ are fixed by the saddle point approximation and are tied to the mean instanton size $\rho$, density $N/V$,  and   determinantal mass $m^*$~\cite{Schafer:1995pz,Faccioli:2001ug,Shuryak:2021fsu,Liu:2024rdm}

The topological screened parameter $\delta$ is fixed to \cite{Liu:2024rdm,Diakonov:1995qy}
\begin{equation}
\delta=N_f\frac{m^*}{m}\frac{\Delta }{N}
\end{equation}
In this saddle point approximation, the constituent mass naturally emerges as 

\begin{equation}
\label{constituent_mass}
    M(k)\simeq \frac N{2N_cV} \frac{k^2\varphi'(k)^2}{m^*}
\end{equation}
which coincides with \eqref{eq:cons_m} by identifying $m^*\approx\sigma_0$ \textcolor{black}{in chiral limit}. At low momenta $k\rho\ll 1$, the  dynamical mass $M(k)$ is about constant $M=M(0)$. At high momenta, the dynamical constituent mass asymptotes the current mass $m$. 

For a canonical ensemble of pseudoparticles, the instanton number sum $N$ and difference $\Delta$ are fixed to  $N=Vn_{I+A}$ and $\Delta=0$, respectively.  In a  grand canonical ensemble, the instanton number sum and difference are allowed to fluctuate.

\subsection{Determinantal mass}
\label{sec:mdet}
The concept of the determinantal mass $m^*$ first emerge as the ensemble average of the emergent vertices $\Theta_I$,

\begin{equation}
\label{eq:det}
    \left\langle 
    \prod_f\rho^{N}\mathrm{Det}(\slashed{D})_{\mathrm{ZM}}\right\rangle\simeq\left\langle\rho^{NN_f} \prod_I\Theta_I\right\rangle=(\rho m^*)^{NN_f}
\end{equation}
which is equivalent to the zero modes estimation for the fermionic determinant in the multi-instanton background. Their values tell us how much the
presence of fermions reduces the instanton density, compared to the same ensemble without them. If one only consider the contributions from zero modes and the lowest order in the quark mass $m$ expansion in the chiral limit, using the effective Lagrangian (\ref{ZNEFF}), one can further estimate the determinantal mass in \eqref{eq:det} in the leading order of $1/N_c$ expansion by computing the VEVs of the instanton determinantal vertices $\langle\theta_\pm\rangle$.  
The same gap equation as in \eqref{eq:det_gap} for the determinantal mass naturally emerges \cite{Liu:2024rdm,Schafer:1996wv},
\begin{equation}
\label{mdet_gap_1}
    m^*\simeq m-\frac{2\pi^2\rho^2}{N_c}\langle \bar \psi_f\mathcal{F}(\rho\partial)\psi_f \rangle
\end{equation}

Using quark current mass $m=6$ MeV, $n_{I+A}=1$ fm$^{-4}$ and $\rho=0.313$ fm. This gives $m^*=103.6$ MeV, which is consistent with the 
result $m^*\sim 103\,\rm  MeV$~\cite{Faccioli:2001ug}, following from numerical simulations of ensembles of interacting pseudoparticles.

In the ILM, the vacuum transition amplitude is 

\begin{equation}
    \frac{n_{I+A}}{2}=\int d\rho n(\rho)\prod_{f=1}^{N_f} (m_f^*\rho)
\end{equation}
where $n_{I+A}$ is the instanton density and $n(\rho)$ denotes the quenched instanton size distribution appeared in \eqref{dn_dist}. The effect of the  determinantal mass $m^*$ is to quench the vacuum tunneling. Thus, the numerical value for the effective instanton-quark coupling in \eqref{eq:effective_action} can be estimated by
\begin{equation}
    \frac{n_{I+A}}{2}
    \left(\frac{4\pi^2\rho^2}{m^*}\right)^{N_f}\approx696.7~\mathrm{GeV}^{-2}~(N_f=2)
\end{equation}




\subsection{Instanton molecules}
\label{sec:mol}
\begin{figure}
    \centering
\subfloat[\label{mol_1}]{\includegraphics[width=0.33\linewidth]{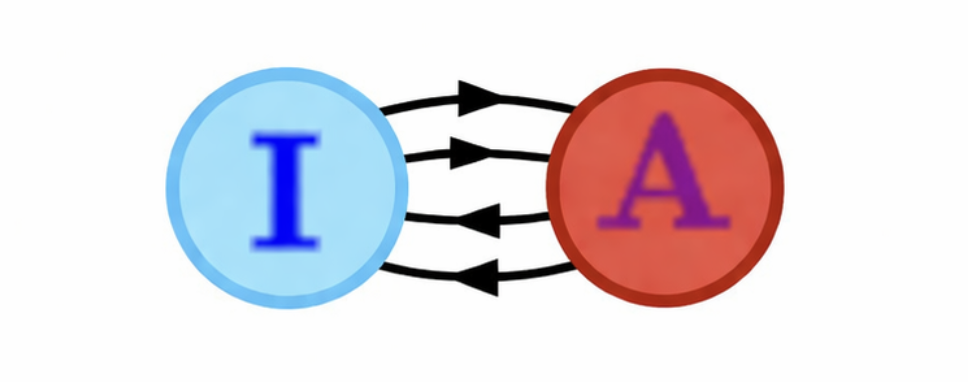}}
\hfill
\subfloat[\label{mol_5}]{\includegraphics[width=0.33\linewidth]{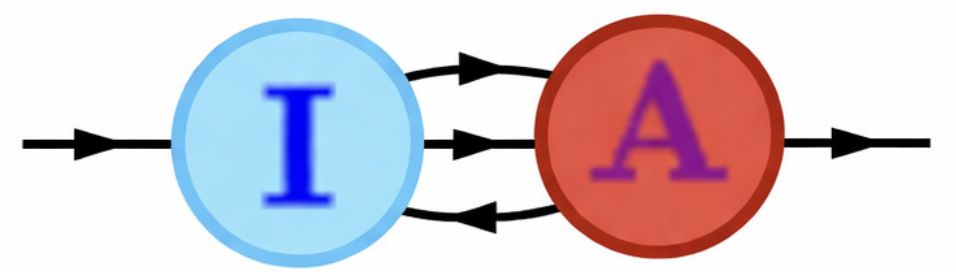}}
\hfill
\subfloat[\label{mol_6}]{\includegraphics[width=0.33\linewidth]{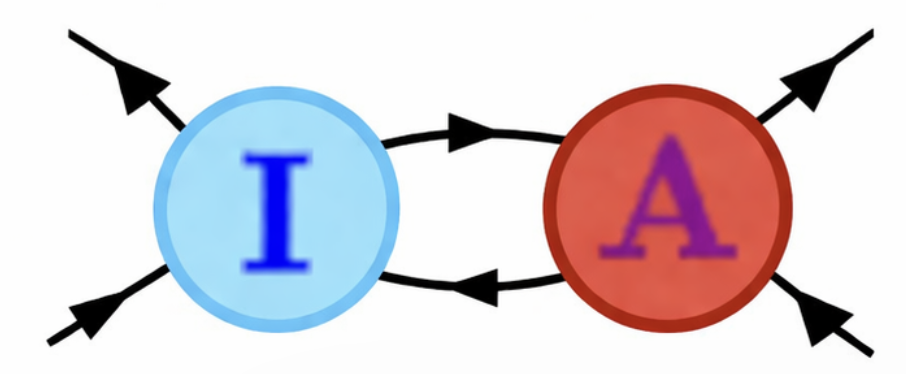}}
    \caption{The Feynman diagrams for the vertices ($N_f=2$) induced by a close pair of instanton ($I$) and anti-instanton ($A$): (a) the vacuum tunneling rate of a fully connected molecule where all flavors looped across the pair. (b) the one-body (two-quark) vertex induced by $\Theta^{(1)}$. (c) the two-body (four-quark) vertex. 
    }
    \label{fig:mole}
\end{figure}

At leading order in density $n_{I+A}$, pseudoparticles in the vacuum are typically randomly distributed. The induced vertices are discussed in Sec.~\ref{sec:tH}. However, when a pair of pseudoparticles comes into close proximity, interactions between them grow significantly, resulting a strongly correlated ensemble where the pair-induced effects are comparable to those from single instantons. To account for these pairs, a systematic cluster expansion that takes into account these correlations among the higher-order density configurations is introduced.

\begin{equation}
\begin{aligned}
   &Z_{N_\pm}\rightarrow
   \int \mathcal{D}\psi \mathcal{D}\psi^\dagger e^{-\int d^4x\left(-\psi^\dagger i\slashed{\partial}\psi+\mathcal{L}_{\rm inst}+\mathcal{L}_{\rm mol}+\cdots\right)}
\end{aligned}
\end{equation}
where the emergent interaction induced by single pseudoparticle reads
\begin{equation}
\begin{aligned}
    \mathcal{L}_{\rm inst}=\int dU\int d\rho n(\rho)\rho^{N_f}\Theta_{I}(z)
\end{aligned}
\end{equation}
and the interaction induced by paired pseudoparticles reads
\begin{equation}
\begin{aligned}
    \mathcal{L}_{\rm mol}=&\int d\rho_Id\rho_An(\rho_I)n(\rho_A)\rho_I^{N_f}\rho_A^{N_f}\\
    &\times\int dud^4R\,\Theta_{IA}(z_I,z_A)
\end{aligned}
\end{equation}
The paired vertices are obtained by connecting quarks between the pair as presented in Fig.~\ref{fig:mole}
\begin{equation}
    \begin{aligned}
\Theta_{IA}=&\left[\Theta_I(z_I)\Theta_A(z_A)\right]_{\rm conn}
    \end{aligned}
\end{equation}

As presented in Fig.~\ref{mol_1}, the vacuum tunneling rate per volume $n_{IA}$ of a pseudoparticle pair can be estimated by 
\begin{equation}
\begin{aligned}
\label{pair_densityIA}
    n_{IA}=&\int d\rho_Id\rho_{A}n(\rho_I)n(\rho_{A})\rho_I^{N_f}\rho_{A}^{N_f}(m^*)^{2N_f}\\
    &\times\int d^4Rdu|T_{IA}(u,R)|^{2N_f}\\    
    =&\left(\frac{n_{I+A}}2\right)^2\int d^4Rdu\left(\frac{|T_{IA}(u,R)|}{m^*}\right)^{2N_f}
\end{aligned}
\end{equation}
where the relative color orientation is $u=U^\dagger_IU_A$ and the separation is defined by $R=z_I-z_A$. 
The first bracket is counting the pair density of pseudoparticles in the vacuum and the second is the \textcolor{black}{effective correlation 4-volume inside the pseudoparticle pair, which is determined by} light quark hopping between the pseudoparticles. 


The $IA$ configurations do not contribute to chiral symmetry breaking due to their chiral preserving nature but their existence are crucial to the helicity preserving bound states such as vector mesons and chirality preserving process. 

The molecular vertices can be obtained by forming the internal quark loops inside the pseudoparticle pair as illustrated in Fig.~\ref{mol_5} and \ref{mol_6}. 
For each flavor passing through a close pair of instanton and anti-instanton as shown in Fig.~\ref{mol_5}, it generates 


\begin{equation}
\begin{aligned}
    &\Theta^{(f)}_{IA}(z_I,z_A)=\frac 1{2N_c}\left(\frac{(4\pi^2\rho^2)^2}{\left|T_{IA}\right|^{2}}\frac{iR_\mu}{R}\frac{dT(R)}{dR}\right)\\
    &\times\bigg[\mathrm{Tr}_c(u\tau_\mu^+ )\mathrm{Tr}_c(\tau^-_\nu u^\dagger)\bar\psi(z_I)\gamma_\nu\frac{1+\gamma^5}{2}\psi(z_A)\\
    &\quad-\mathrm{Tr}_c(\tau_\mu^-u^\dagger)\mathrm{Tr}_c(u\tau^+_\nu)\bar\psi(z_A)\gamma_\nu\frac{1-\gamma^5}{2}\psi(z_I)\bigg]
\end{aligned}
\end{equation}

The molecular vertices are composed of one-body and two-body vertices shared between any pseudoparticle pairs. 
The molecular coupling strength are the combination of the molecule transition rate 
\eqref{pair_densityIA} and quark-pair coupling induced as the quark passing through the close pair. 

In the $1/N_c$ counting for single instanton framework, each quark flavor passing through the pseudoparticle contributes a factor of $1/N_c$, while each quark loop  introduces an additional factor of $N_c$. The instanton density also scales as $N_c$, ensuring that the effective mass $m^*\sim\sqrt{n_{I+A}/N_c}$ remains of order unity. 

\section{Gluon-instanton interactions}
\label{Feyn}

In this section, the interaction between gluons and instantons is discussed. 
The interaction between pseudoparticles with opposite topological charges at large distances was first derived in \cite{Callan:1977gz,Forster:1977jv} by studying the interaction of an instanton with a weak, slowly varying external field strength $F_{\mu\nu}$. The instanton field is put in the singular gauge in order to ensure that the gauge field is localized. With this in mind, one finds

\begin{equation}
\begin{aligned}
\label{inst_dipole}
     S_{int}=&-i\frac{2\pi^2\rho^2}{g}\mathrm{tr}_c\left[U_I\tau^-_\mu\tau^+_\nu U_I^\dagger F_{\mu\nu}\right]\\
     =&\frac{2\pi^2\rho^2}{g}R^{ab}(U_I)\bar\eta^b_{\mu\nu}F^a_{\mu\nu}
\end{aligned}
\end{equation}

This result can be interpreted as a classical external field coupling to the color magnetic dipole moment $\frac{2\pi^2 \rho^2}{g} \bar{\eta}^a_{\mu\nu}$ of the instanton. If the external field is assumed to be an anti-instanton located at a relative distance $R$, Eq.~\eqref{inst_dipole} can be used to describe the interaction between well-separated pseudoparticles with opposite topological charges. Thus, the gauge interaction $S_{int}$ in \eqref{eq:Z_N} can be written as

\begin{equation}
\label{S_int}
    S_{int}=\frac{32\pi^2}{g^2}\rho_I^2\rho_A^2\bar\eta_{\mu\rho}^a\eta_{\nu\rho}^bR^{ab}(U_{IA})\frac{R_\mu R_\nu}{R^6}
\end{equation}

This semi-classical gauge interaction $S_{int}$ between instanton and anti-instanton can be also reproduced by the amplitude for semi-classical color force exchanges between the instantons and anti-instantons \cite{Zakharov:1992bx}
\begin{widetext}
\begin{equation}
\begin{aligned}
    &\left\langle\exp\left(-\frac{2\pi^2\rho^2_I}{g}R^{ab}(U_I)\bar\eta^b_{\mu\nu}F^a_{\mu\nu}\right)\exp\left(-\frac{2\pi^2\rho^2_A}{g}R^{cd}(U_A)\eta^d_{\rho\lambda}F^c_{\rho\lambda}\right)\right\rangle\\
    =&1+\frac{4\pi^4}{g^2}\rho_I^2\rho_A^2R^{ab}(U_I)R^{cd}(U_A)\bar\eta_{\mu\nu}^b\eta_{\rho\lambda}^d\left\langle F^a_{\mu\nu}(z_I)F^c_{\rho\lambda}(z_J)\right\rangle+\cdots\\
    =&e^{-S_{int}}
\end{aligned}
\end{equation}
by summing all color force exchanges with the given free propagators, 
\begin{equation}
    \langle F^a_{\mu\nu}(x)F^b_{\rho\lambda}(0)\rangle=-\frac{2\delta^{ab}}{\pi^2x^6}\left[x_\mu x_\rho\delta_{\nu\lambda}-x_\mu x_\lambda\delta_{\nu\rho}-x_\nu x_\rho\delta_{\mu\lambda}+x_\nu x_\lambda\delta_{\mu\rho}-\frac{x^2}2\left(\delta_{\mu\rho}\delta_{\nu\lambda}-\delta_{\mu\lambda}\delta_{\nu\rho}\right)\right]
\end{equation} 
\end{widetext}
This color force is formed by overlapping the tails of each semiclassical profiles of pseudoparticles. Since well-separated instanton–anti-instanton pairs are not significantly distorted, their interaction is well-defined semi-classically. For very close pairs, on the other hand, the instanton fields are strongly distorted. On top of that, the perturbative feature, which occurs when the close instanton-anti-instanton pairs begin to annihilate, is not included in semi-classical approximations. Thus, both the strong distortion and the perturbative feature leave the description of the interaction uncertain.

\textcolor{black}{With the presence of instantons in the vacuum, one can also calculate the gluon emission from the vacuum semiclassically by well-separated pseudoparticles. The single-instanton vacuum transition amplitude with one gluon emission is sourced by a semi-classical instanton profile at $z_I$} 

\begin{equation}
\begin{aligned}
\label{LSZ_g}
    &A_{I}(z_I)\\
    &=\int \frac{d^4k}{(2\pi)^4}e^{ik\cdot z_I} \left[-i\frac{4\pi^2\rho^2}g\bar\eta^a_{\mu\nu}\frac{k_\nu}{k^2}\mathcal{F}_g(\rho k)\right]\\
    &=\frac{2\pi^2\rho^2}{g}\bar{\eta}_{\mu\nu}^a\mathcal{F}_g(i\rho\partial_{z_I}) \langle k| F^a_{\mu\nu}(z_I) |0\rangle_{\mathrm{free}}\\
    &\xrightarrow{\rho\rightarrow0}\frac{2\pi^2\rho^2}{g}\bar{\eta}_{\mu\nu}^a\langle k| F^a_{\mu\nu}(z_I)|0\rangle_{\mathrm{free}}
\end{aligned}
\end{equation}
\textcolor{black}{where $F^a_{\mu\nu}(x)$ is the gluon field strength carrying free plane wave modes and $|k\rangle$ is the single gluon Fock state that projects out the emitted gluon.}


This allows us to include the gluon degrees of freedom as semiclassical gauge interactions in our effective instanton liquid Lagrangian in \eqref{eq:Z_N2}. The linearized interaction \eqref{inst_dipole} is attached to the instanton vertex $\Theta_{I}(x)$ in \eqref{eq:tHooft} exponentially, resulting in \cite{Vainshtein:1981wh}

\begin{align}
\label{eq:tHooft_g}
\Theta_{I}(x)\rightarrow \Theta_{I}(x)e^{i\frac{2\pi^2\rho^2}{g}\mathrm{tr}_c\left[U_I\tau^\mp_\mu\tau^\pm_\nu U_I^\dagger F_{\mu\nu}\right]} 
\end{align}
The finite size effect of instanton color-magnetic moment can be straightforwardly recovered by 
\begin{align}
\frac{2\pi^2\rho^2}{g}\rightarrow \frac{2\pi^2\rho^2}{g}\mathcal{F}_g(\rho q)
\end{align}
where the finite-sized color-magnetic moment profile is defined as
\begin{equation}
\label{eq:g_form}
    \mathcal{F}_g(q)=\frac{4}{q^2}-2K_2(q)
\end{equation}

The color field strength $F_{\mu\nu}$ follows from the LSZ reduction of pseudoparticle field profile, and is coupled to the color-magnetic moment of individual instantons~\cite{Kochelev:1996pv,Qian:2015wyq,Diakonov:2002fq}. It can be interpreted as color field sourcing the tail of the instanton profiles.

\begin{figure}
\centering
\subfloat[\label{fig:tHooft_g}]{\includegraphics[width=.99\linewidth]{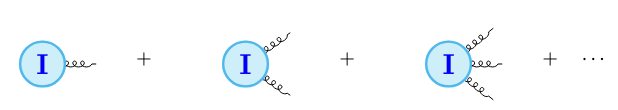}}
\hfill
\subfloat[\label{fig:tHooft_qg1}]{\includegraphics[width=.3\linewidth]{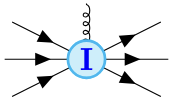}}
\hfill
\subfloat[\label{fig:tHooft_qg2}]{\includegraphics[width=.3\linewidth]{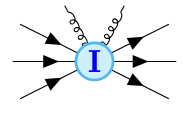}}
\caption{Additional Feynmann diagrams in the ILM after including gluon emission. (a) semi-classical emission of gluon plane wave from instanton vacuum. (b), (c) quark 't Hooft vertices combined with semi-classical gluon emission }
\label{fig:ILM}
\end{figure}




The effective Lagrangian $\mathcal{L}_{\rm eff}^{q+g}$ following from (\ref{eq:tHooft_g}) after averaging over the color orientation yield emergent multi-flavor interactions with instanton tails sourced by external color fields. 
The full quark and gluon vertices are graphically presented in Fig.~\ref{fig:ILM}. 

When the color sourcing field $F_{\mu\nu}$ contracts with the outgoing gluon states, it can also be viewed as gluon emission from the fail vacuum tunneling as illustrated in Fig.~\ref{fig:tHooft_g}. Another important feature introduced by this gluon-instanton vertex is the anomalous quark chromomagnetic moment \cite{Kochelev:1996pv} as illustrated in Fig.~\ref{fig:tHooft_qg1}, which has significant applications, including the study of gluon distributions in hadrons \cite{Kochelev:2015pqd}, the odderon \cite{Kochelev:2013csa}, the Pauli form factor \cite{Zhang:2017zpi}, and spin physics \cite{Cherednikov:2006zn,Kochelev:2008fy}. Furthermore, when the color sourcing field $F_{\mu\nu}$ contracts with the QCD operators, it reproduces the semi-classical field insertions in the multi-instanton expansion of the operators \eqref{eq:gluo_op} with the small $\rho$ expansion. Thus, the instanton insertion now can be rewritten as gluon-instanton vertices by introducing a color source located at each instanton center $z_I$. \eqref{eq:inst_contribution} and \eqref{eq:o_had_exp} can simply be presented by the path integral with full Lagrangian $\mathcal{L}^{q+g}_{\rm eff}$ obtained by vertices in \eqref{eq:tHooft_g}.

These interactions in $\mathcal{L}^{q+g}_{\rm eff}$ have similar book-keeping in $1/N_c$. For a single instanton with $n_q$ open quark flavors and $n_g$ gluons, the vertices in \eqref{eq:tHooft} and \eqref{eq:tHooft_g} give rise to an effective 't Hooft interaction coupling. 
$$
G_I^{q+g}\sim\frac{n_{I+A}}{2}\frac{1}{N_c^{n_q+n_g}}\left(\frac{4\pi^2\rho^2}{m^*}\right)^{n_q}\left(\frac{2\pi^2\rho^2}{g}\right)^{n_g}
$$
The emergent couplings with constituent quarks and gluons are determined through color averaging, with each $UU^\dagger$ pair contributing a $1/N_c$ factor in the large $N_c$ limit.
Note that each gluon emission from the instanton is further suppressed by the instanton size. Now the effective interactions are given by \cite{Liu:2024rdm}.

\begin{widetext}
\begin{equation}
\begin{aligned}
\label{LEFFSIA}
\mathcal{L}^{q+g}_{\mathrm{eff}}=&\bar{\psi}(i\slashed{\partial}-m)\psi+\frac{n_{I+A}}{2}\mathcal{V}_{n_q=0}+\frac{n_{I+A}}{2}\left(\frac{4\pi^2\rho^2}{m^*}\right)\mathcal{V}_{n_q=1}+\frac{n_{I+A}}{2}\left(\frac{4\pi^2\rho^2}{m^*}\right)^2\mathcal{V}_{n_q=2}+\mathcal{O}\left(\frac{n_{I+A}}{2}\left(\frac{4\pi^2\rho^2}{m^*}\right)^3\right)
\end{aligned}
\end{equation}
where the zero-body (gluodynamic) interaction is defined as
\begin{equation}
    \mathcal{V}_{n_q=0}=\frac{1}{N^2_c-1}\left(\frac{2\pi^2\rho^2}{g}\right)^2(F^a_{\mu\nu})^2+\frac{4}{3N_c(N^2_c-1)}\left(\frac{2\pi^2\rho^2}{g}\right)^3f^{abc}F^a_{\mu\nu}F^b_{\mu\lambda}F^c_{\nu\lambda}+\mathcal{O}\left(\left(\frac{2\pi^2\rho^2}{g}\right)^4\right)
\end{equation}
and the one-body interaction is defined as
\begin{equation}
\begin{aligned}
   \mathcal{V}_{n_q=1}=& -\frac{1}{N_c}\bar{\psi}\psi+\frac{1}{N^2_c-1}\left(\frac{2\pi^2\rho^2}{g}\right)\bar{\psi}\sigma_{\mu\nu}\frac{\lambda^a}2\psi F^a_{\mu\nu}-\frac{1}{N_c(N^2_c-1)}\left(\frac{2\pi^2\rho^2}{g}\right)^2f^{abc}\bar{\psi}\sigma_{\mu\nu}\lambda^a\psi F^b_{\mu\rho}F^c_{\nu\rho}\\
&-\frac{1}{N_c(N^2_c-1)}\left(\frac{2\pi^2\rho^2}{g}\right)^2\left(\delta^{bc}\bar{\psi}\psi+\frac{N_c}{4(N_c+2)}d^{abc}\bar{\psi}\lambda^a\psi\right) F^b_{\mu\nu}F^c_{\mu\nu}\\
&-\frac{1}{N_c(N^2_c-1)}\left(\frac{2\pi^2\rho^2}{g}\right)^2\left(\delta^{bc}\bar{\psi}\gamma^5\psi+\frac{N_c}{4(N_c+2)}d^{abc}\bar{\psi}\lambda^a\gamma^5\psi\right)F^b_{\mu\nu}\tilde{F}^c_{\mu\nu}\\
&+\mathcal{O}\left(\left(\frac{2\pi^2\rho^2}{g}\right)^3\right)\\
\end{aligned}
\end{equation}
and the two-body interaction is defined as
\begin{equation}
\begin{aligned}
    \mathcal{V}_{n_q=2}=&\frac{2N_c-1}{16N_c(N^2_c-1)}\left[(\bar{\psi}\psi)^2-(\bar{\psi}\tau^a\psi)^2-(\bar{\psi}i\gamma^5\psi)^2+(\bar{\psi}i\gamma^5\tau^a\psi)^2\right]\\
    &+\frac{1}{32N_c(N^2_c-1)}\left[\left(\bar{\psi}\sigma_{\mu\nu}\psi\right)^2-\left(\bar{\psi}\sigma_{\mu\nu}\tau^a\psi\right)^2\right]\\
    &-\frac{1}{N_c(N_c^2-1)}\left(\frac{2\pi^2\rho^2}{g}\right)\left[\bar{u}_Ru_L\bar{d}_R\sigma_{\mu\nu}\frac{\lambda^a}{2}d_L+\bar{u}_R\sigma_{\mu\nu}\frac{\lambda^a}{2}u_L\bar{d}_Rd_L\right]F^a_{\mu\nu}\\
    &-\frac{1}{(N_c+2)(N_c^2-1)}\left(\frac{2\pi^2\rho^2}{g}\right)d^{abc}\left[\bar{u}_R\frac{\lambda^a}{2}u_L\bar{d}_R\sigma_{\mu\nu}\frac{\lambda^b}{2}d_L+\bar{u}_R\sigma_{\mu\nu}\frac{\lambda^a}{2}u_L\bar{d}_R\frac{\lambda^b}{2}d_L\right]F^c_{\mu\nu}\\
    &-\frac{1}{(2N_c)(N_c^2-1)}\left(\frac{2\pi^2\rho^2}{g}\right)f^{abc}\left[\bar{u}_R\sigma_{\mu\rho}\frac{\lambda^a}{2}u_L\bar{d}_R\sigma_{\nu\rho}\frac{\lambda^b}{2}d_L+\bar{u}_R\sigma_{\mu\rho}\frac{\lambda^a}{2}u_L\bar{d}_R\sigma_{\nu\rho}\frac{\lambda^b}{2}d_L\right](F^c_{\mu\nu}-\tilde{F}^c_{\mu\nu})\\
    &+\mathcal{O}\left(\left(\frac{2\pi^2\rho^2}{g}\right)^2\right)\\
\end{aligned}
\end{equation}
\end{widetext}
Here we present pure color sources, one-body interactions with up to three semi-classical gluons, and two-body interactions with one color sourcing field. Higher-order interactions follows similar scaling but increases in complexity. By switching off the color sourcing field strength, the effective Lagrangian reduces back to 't-Hooft Lagrangian \eqref{THOOFT1} of any flavor number. The $1/N_c$ counting of each vertex 
implies the more quarks and gluons involved in the instanton, the more $1/N_c$ suppression.

The use of the gluonic vertices in \eqref{eq:tHooft_g} is justified in momentum space diagrams, when the exchanging semi-classical gluons carry energies below the sphaleron mass
(the top of the tunneling barrier)
\begin{equation}
M_S=\int d^3x \frac 18{F^2_{\mu\nu}(0,\vec x)}
=\frac{3\pi}{4\alpha_s\rho}
\end{equation}
With $8\pi^2/g^2(\rho)=10$--$15$~\cite{Schafer:1996wv}, The sphaleron mass is given by $M_S\sim 2.5$ GeV, for $\alpha_s(1/\rho)\sim 0.42$--$0.7$.


\section{Instanton-based effective theories for quarks}
\label{effective_Langrangian}
\begin{figure}
    \centering
    \includegraphics[width=1\linewidth]{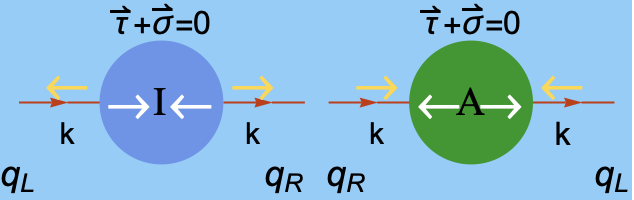}
    \caption{Light quarks flip their chirality when passing through the instanton (left) and anti-instanton (right)}
    \label{fig:inst-q}
\end{figure}

At low resolution, the QCD vacuum is predominantly populated by topologically active instantons and anti-instantons, which are Euclidean tunneling configurations between vacua with different winding numbers. In Fig.~\ref{fig:inst-q}, light quarks interacting with these topological configurations develop zero modes with fixed handedness. For instance, a massless left-handed quark tunneling through an instanton can appear as a right-handed massless quark. The same scenario happens at an anti-instanton with the handedness of the quark flipped.

For a single quark flavor, this mechanism is at the origin of the explicit breaking of $U_A(1)$ symmetry.
For multiple light quark flavor, this mechanism can account for the dual breaking of the $U_A(1)$  (explicitly)  and chiral symmetry (spontaneously). This is manifested through the emergent multi-flavored interactions induced by the light quark zero modes.

\subsection{'t Hooft Lagrangian}
\label{tHooft_L}
In the instanton vacuum, these multi-flavored interactions are the well-known $^\prime$t Hooft determinantal interactions.
In the local approximation, the instanton size is taking to zero. 

\subsubsection{Two-flavor case}

By explicitly carrying out the color average in the effective Lagrangian in \eqref{eq:effective_action} for $N_f=2$, the induced interactions for the light quarks from single instanton plus anti-instanton give \cite{Liu:2023fpj,Rapp:1999qa,Shuryak:2021yif}
\begin{widetext}
\begin{equation}
\label{THOOFT1}
\begin{aligned}
    \mathcal{L}_I=&\frac{G_I}{8(N^2_c-1)}\left\{\frac{2N_c-1}{2N_c}\left[(\bar{\psi}\psi)^2-(\bar{\psi}\tau^a\psi)^2-(\bar{\psi}i\gamma^5\psi)^2+(\bar{\psi}i\gamma^5\tau^a\psi)^2\right]-\frac{1}{4N_c}\left[\left(\bar{\psi}\sigma_{\mu\nu}\psi\right)^2-\left(\bar{\psi}\sigma_{\mu\nu}\tau^a\psi\right)^2\right]\right\}\\
\end{aligned}
\end{equation}
\end{widetext}
which are seen to mix $LR$ chiralities. The  effective coupling 
\begin{equation}
    G_I=\int d\rho n(\rho)\rho^{N_f}(4\pi^2\rho^2)^{N_f}\simeq\frac{n_{I+A}}{2}
    \left(\frac{4\pi^2\rho^2}{m^*}\right)^{N_f}
\end{equation}
is fixed by the mean-instanton density and $m^*_f$ the induced determinantal mass~\cite{Vainshtein:1981wh}. Note that there is no vector or axial vector channel in \eqref{THOOFT1}. To obtain vector bound states, it is necessary to go beyond the single instanton-induced interaction. In the interacting instanton vacuum, additional multi-flavor interactions involving clusted instantons are expected, as discussed in Sec.~\ref{sec:mol}. Given the diluteness of the tunneling processes in the QCD vacuum at low resolution, the natural interactions are molecular in the form of binary instanton-anti-instanton configurations. When the relative orientation is maximally locked in color space, they induce flavor mixing interactions of the form~\cite{Liu:2023fpj,Rapp:1999qa,Schafer:1994nv}
\begin{widetext}
\begin{equation}
\label{THOOFT2}
 \begin{aligned}
\mathcal{L}_{IA}=G_{IA}\bigg\{&\frac{1}{N_c(N_c-1)}\left[(\bar{\psi}\gamma^\mu\psi)^2+(\bar{\psi}\gamma^\mu\gamma^5\psi)^2\right]-\frac{N_c-2}{N_c(N_c^2-1)}\left[(\bar{\psi}\gamma^\mu\psi)^2- (\bar{\psi}\gamma^\mu\gamma^5\psi)^2\right]\\
    &+\frac{2N_c-1}{N_c(N_c^2-1)}\left[(\bar{\psi}\psi)^2+(\bar{\psi}\tau^a\psi)^2+(\bar{\psi}i\gamma^5\psi)^2+(\bar{\psi}i\gamma^5\tau^a\psi)^2\right]\\
     &-\frac{1}{2N_c(N_c-1)}\left[(\bar{\psi}\gamma^\mu\psi)^2+(\bar{\psi}\tau^a\gamma^\mu\psi)^2+(\bar{\psi}\gamma^\mu\gamma^5\psi)^2+(\bar{\psi}\tau^a\gamma^\mu\gamma^5\psi)^2\right]\bigg\}  
\end{aligned}
\end{equation}
which are $LL$ and $RR$ chirality preserving, in contrast to (\ref{THOOFT1}). The effective molecule-induced coupling is defined as
 \begin{equation}
 \begin{aligned}
 \label{MOLX}
     G_{IA}=&\int d\rho_I d\rho_{A}\int dud^4R ~ \frac1{8}(4\pi^2\rho^2_I)(4\pi^2\rho^2_{A})n(\rho_I)n(\rho_{A})\rho_I^{N_f}\rho_{A}^{N_f}|T_{IA}(u,R)|^{2N_f-2}\\
\end{aligned}
\end{equation}
 \end{widetext}
Here $R=z_I-z_{A}$ is the relative molecular separation, $u=U_{A} U^\dagger_I$ is the relative
molecular color orientation, and $T_{IA}$ is the hopping quark matrix jumping from anti-instanton to instanton and its conjugate defines the reverse process.

\begin{equation}
\begin{aligned}
\label{TIA}
    T_{IA}(u,R)&=\int d^4x\phi_I^\dagger(x-z_I) i\slashed{\partial}\phi_A(x-z_A)\\
\end{aligned}
\end{equation}

The $T_{IA}^{-2}$ in the molecule-induced coupling in (\ref{MOLX}) is readily understood that a pair of quark lines is removed in Fig.~\ref{mol_1} by the division of $|T_{IA}|^2$.
The strength of the induced molecular coupling $G_{IA}$ to the single coupling $G_I$ can be parameterized as

\begin{equation}
G_{IA}=\frac18G_I^2\int dud^4R\left(\frac{T_{IA}(u,R)}{4\pi^2\rho^2}\right)^{2N_f-2}
\end{equation}
In $1/N_c$ limit, the vacuum expectation value of the hopping between the pair is estimated
\begin{equation}
\begin{aligned}
\label{eq:hop_parameter}
    &\int dud^4R[\rho T_{IA}(u,R)]^{2N_f}\\
   &\simeq N_f!\left(\frac{2}{N_c}\right)^{N_f}\rho^4\int d^4R(4\pi^2\rho^3T'(R))^{2N_f}
\end{aligned}
\end{equation}



Low-lying meson dynamics at the low energy can be completely described by the instanton-induced interaction \cite{Liu:2023fpj,Shuryak:2021fsu}. To have physical mass spectrum of light mesons consistent with the experiments, the parameters in the 't Hooft Lagrangian has to be fixed by specific values. 
These values are subject to the values of instanton size $\rho$ and instanton density $n_{I+A}$. 


\subsubsection{Three-flavor case}
The same color average in \eqref{eq:effective_action} for $N_f=3$ carries out the induced interactions for the light quarks $u,d,s$. The interaction reads
\begin{widetext}

\begin{align}
\mathcal{L}^{N_f=3}_{I}
&= -  
\frac{G_I}{N_c (N_c^2 - 1)}
\left(
\frac{2N_c + 1}{2N_c + 4}\mathrm{det}\bar{\psi}_R\psi_L
+ \frac{3}{8 (N_c + 1)}\mathrm{det}\bar{\psi}_R\sigma_{\mu\nu}\psi_L
\right)
+ (L \leftrightarrow R)
\end{align}
where the typical 't Hooft determinantal interaction in $N_f=3$ is defined as,

\begin{equation}
\mathrm{det}\bar{\psi}_R\psi_L=\frac16\epsilon_{ijk}\epsilon_{lmn}\bar{\psi}_{R i}\psi_{L l}\,
\bar{\psi}_{R j}\psi_{L m}\,
\bar{\psi}_{R k}\psi_{L n}=\left|\begin{array}{ccc}
\bar{u}_Ru_L & \bar{u}_Rd_L & \bar{u}_Rs_L \\
\bar{d}_Ru_L & \bar{d}_Rd_L & \bar{d}_Rs_L \\
\bar{s}_Ru_L & \bar{s}_Rd_L & \bar{s}_Rs_L \end{array}\right|
\end{equation}
and
\begin{equation}
\begin{aligned}
&\mathrm{det}\bar{\psi}_R\sigma_{\mu\nu}\psi_L=   \frac16\epsilon_{ijk}\epsilon_{lmn} \,
\bar{\psi}_{R i}\psi_{L l}\,
\bar{\psi}_{R j}\sigma_{\mu\nu}\psi_{L m}\,
\bar{\psi}_{R k}\sigma_{\mu\nu}\psi_{L n}\\
&=\bar u_{R}u_{L}\left|\begin{array}{cc}
\bar d_{R}\sigma_{\mu\nu}d_{L} & \bar d_{R}\sigma_{\mu\nu}s_{L} \\
\bar s_{R}\sigma_{\mu\nu}d_{L} & \bar s_{R}\sigma_{\mu\nu}s_{L}  
\end{array}\right|
-\bar u_{R}d_{L}\left|\begin{array}{cc}
\bar d_{R}\sigma_{\mu\nu}u_{L} & \bar d_{R}\sigma_{\mu\nu}s_{L} \\
\bar s_{R}\sigma_{\mu\nu}u_{L} & \bar s_{R}\sigma_{\mu\nu}s_{L}  
\end{array}\right|
+\bar u_{R}s_{L}\left|\begin{array}{cc}
\bar d_{R}\sigma_{\mu\nu}u_{L} & \bar d_{R}\sigma_{\mu\nu}d_{L} \\
\bar s_{R}\sigma_{\mu\nu}u_{L} & \bar s_{R}\sigma_{\mu\nu}d_{L}  
\end{array}\right|\\
&-\bar d_{R}u_{L}\left|\begin{array}{cc}
\bar u_{R}\sigma_{\mu\nu}d_{L} & \bar u_{R}\sigma_{\mu\nu}s_{L} \\
\bar s_{R}\sigma_{\mu\nu}d_{L} & \bar s_{R}\sigma_{\mu\nu}s_{L}  
\end{array}\right| +\bar d_{R}d_{L}\left|\begin{array}{cc}
\bar u_{R}\sigma_{\mu\nu}u_{L} & \bar u_{R}\sigma_{\mu\nu}s_{L} \\
\bar s_{R}\sigma_{\mu\nu}u_{L} & \bar s_{R}\sigma_{\mu\nu}s_{L}  
\end{array}\right|
-\bar d_{R}s_{L}\left|\begin{array}{cc}
\bar u_{R}\sigma_{\mu\nu}u_{L} & \bar u_{R}\sigma_{\mu\nu}d_{L} \\
\bar s_{R}\sigma_{\mu\nu}u_{L} & \bar s_{R}\sigma_{\mu\nu}d_{L}  
\end{array}\right|\\
&+\bar s_{R}u_{L}\left|\begin{array}{cc}
\bar u_{R}\sigma_{\mu\nu}d_{L} & \bar u_{R}\sigma_{\mu\nu}s_{L} \\
\bar d_{R}\sigma_{\mu\nu}d_{L} & \bar d_{R}\sigma_{\mu\nu}s_{L}  
\end{array}\right|
-\bar s_{R}d_{L}\left|\begin{array}{cc}
\bar u_{R}\sigma_{\mu\nu}u_{L} & \bar u_{R}\sigma_{\mu\nu}s_{L} \\
\bar d_{R}\sigma_{\mu\nu}u_{L} & \bar d_{R}\sigma_{\mu\nu}s_{L}  
\end{array}\right|+\bar s_{R}s_{L}\left|\begin{array}{cc}
\bar u_{R}\sigma_{\mu\nu}u_{L} & \bar u_{R}\sigma_{\mu\nu}d_{L} \\
\bar d_{R}\sigma_{\mu\nu}u_{L} & \bar d_{R}\sigma_{\mu\nu}d_{L}  
\end{array}\right|
\end{aligned}
\end{equation}

\end{widetext}
\subsection{Bosonization}

By averaging over the color orientation of instantons using $1/N_c$ as a book-keeping argument, the leading order of the 't Hooft effective Lagrangian in \eqref{eq:effective_action} reads

\begin{equation}
\label{tHooft}
\mathcal{L}_{\mathrm{eff}}=\bar{\psi}\left(i\slashed{\partial}-m\right)\psi-\frac{G_I}{N_c^{N_f}}\left(\mathrm{det}\bar{\psi}_L\psi_R+\mathrm{det}\bar{\psi}_R\psi_L\right)
\end{equation}


This framework classifies QCD degrees of freedom into two categories: (i) those with masses \(\geq 1/\rho\) and (ii) those with masses \(\ll 1/\rho\). In low-energy strong interactions, where momenta are much smaller than \(1/\rho \simeq 600\) MeV, the heavy modes can be neglected, focusing only on the light degrees of freedom. The only relevant low-mass states are Goldstone pseudoscalar mesons and quarks, which acquire a dynamically generated mass \(M \simeq 300\)–\(400\) MeV \(\ll 1/\rho\). Consequently, for momenta \(k \ll 1/\rho\), QCD reduces to a simpler yet nontrivial theory of massive quarks interacting with nearly massless pions. With this in mind, at lower energy scale $\mu\sim300-400$ MeV, the Lagrangian in \eqref{tHooft} can be approximately bosonized by introducing \(N_f \times N_f\) auxiliary fields, as detailed in \cite{Kacir:1996qn}.

\begin{widetext}
\begin{equation}
\label{semi-bos}
    \mathcal{L}_{bos}=\bar{\psi}(i\slashed{\partial}-m)\psi+\frac{2\pi^2\rho^2}{N_c}\bar{\sigma}\bar\psi\left[\frac{1-\gamma^5}{2}U+\frac{1+\gamma^5}{2}U^\dagger\right]\psi+\frac12\mathrm{Tr}\left[m\bar{\sigma}(U+U^\dagger)\right]
\end{equation}
\end{widetext}
where $N_f\times N_f$ auxiliary bosonic field is defined as
\begin{equation}
U=\exp\left(i\pi^a\tau^a/F_\pi\right)
\end{equation}

The second term in \eqref{semi-bos} represents the quark-meson effective interaction with Goldberger-Treiman (GT) relation manifested.

\begin{equation}
    g_{\pi qq}=\frac{2\pi^2\rho^2}{N_c}\frac{\bar\sigma}{F_\pi}=\frac{M}{F_\pi}
\end{equation}
where $\bar\sigma=-\langle\bar q q\rangle$ with the identification of the constituent mass by \eqref{constituent_mass} and quark condensate in \eqref{qq}. The last term determines the mass of the (pseudo) Goldstone boson by Gell-Mann-Oakes-Renner (GOR) relation.

\begin{equation}
    m^2_\pi=\frac{2m\bar\sigma}{F_\pi^2}
\end{equation}

More specifically, the semi-bosonized Lagrangian in \eqref{semi-bos} can be rewritten as \cite{Diakonov:1995ea}

\begin{equation}
\label{chiral}
    \mathcal{L}_{bos}=\bar{\psi}(i\slashed{\partial}-MU^{\gamma^5})\psi
\end{equation}
by using the identity

\begin{equation}
    \frac{1-\gamma^5}{2}U+\frac{1+\gamma^5}{2}U^\dagger=U^{\gamma^5}
\end{equation}
where the pseudoscalar Goldstone modes are manifested by
\begin{equation}
U^{\gamma^5}=\exp\left(i\pi^a\tau^a\gamma^5/F_\pi\right)
\end{equation}

\subsubsection{Chiral Lagrangian}

If one integrates off the quark fields in \eqref{chiral}, one gets the effective chiral
Lagrangian. This idea also is also known as chiral-quark-soliton model \cite{Diakonov:1997sj}.
\begin{equation}
\begin{aligned}
\label{ChPT}
&S_{\mathrm{\chi PT}}=\frac{F_\pi^2}{4}\int d^4x\Tr\left(L_\mu L_\mu\right)\\
&-\frac{N_c}{192\pi^2}\int d^4x\left[2\Tr(\partial_\mu L_\mu)^2+\Tr (L_\mu L_\nu L_\mu L_\nu)\right]\\
&+\frac{N_c}{240\pi^2}\int d^5x\epsilon_{\mu\nu\rho\lambda\sigma}\Tr(L_\mu L_\nu L_\rho L_\lambda L_\sigma)
\end{aligned}
\end{equation}
where the chiral field is defined as
\begin{align}
    L_\mu=iU^\dagger\partial_\mu U
\end{align}

The first term here is the old Weinberg chiral lagrangian \cite{Weinberg:1991um} with
\begin{equation}
    F^2_\pi=4N_c\int \frac{d^4k}{(2\pi)^4}\frac{M^2(k)}{[k^2+M^2(k)]^2}
\end{equation}
which has also been observed in light-front formulation of instanton model \cite{Liu:2021evw}. The second term are the four-derivative Gasser–Leutwyler terms \cite{Gasser:1983yg,Scherer:2002tk} (with coefficients which turn out to agree with those following from the analysis of the data); the last term is the so-called Wess–Zumino term \cite{Lee:2020ojw}. Note that the $F_\pi$ constant diverges logarithmically at large momenta but is smoothly cut by the momentum dependent mass at $k \sim 1/\rho$ as a result of the finite instanton size.

\section{QCD operators in instanton ensemble}
\label{operators}
This section presents a general framework to calculate the VEVs, hadron matrix elements, and hadronic form factors in the instanton liquid model. Let ${\cal O}_{\mathrm{QCD}}$ be a generic QCD operator where the gluonic part is sourced by a multi-pseudoparticle gluon field given by the sum ansatz at low resolution~\cite{Diakonov:1995qy}

\begin{equation}
\label{eq:gluon_field}
    A_{\mathrm{inst}}(x)=\sum_{I=1}^{N_++N_-} A_I(x)
\end{equation} 

The ensuing gluonic operator ${\cal O}_{\mathrm{QCD}}[\psi,\bar\psi,A]$ splits
into a sum of operators with multi-instanton sources
\begin{equation}
\begin{aligned}
\label{eq:gluo_op}
    \mathcal{O}_{\mathrm{QCD}}=&\sum_{I}\mathcal{O}[\psi,\bar\psi,A_I]\\
    &+\sum_{I\neq J}\mathcal{O}[\psi,\bar\psi,A_I,A_J]+\cdots
\end{aligned}
\end{equation}
of increasing complexity. Typically, the gluonic field strength for the multi-instanton configuration can be split into single instanton fields, and crossing terms typical for non-Abelian fields,
\begin{equation}
\label{eq:field_strength}
F_{\mu\nu}[A_{\mathrm{inst}}]=\sum_{I}F_{\mu\nu}[A_I]+\sum_{I\neq J}F_{\mu\nu}[A_{I},A_{J}]
\end{equation}





\subsection{Vacuum operator expansion}
The vacuum averages of local QCD operators using (\ref{eq:Z_N2}) for fixed-$N_\pm$ configurations (canonical ensemble) are given by

\begin{equation}
\label{eq:op_N}
    \begin{aligned}
        \langle\mathcal{O}_{\mathrm{QCD}}\rangle_{N_\pm}=&\int \mathcal{D}\psi \mathcal{D}\psi^\dagger e^{\int d^4x \psi^\dagger i\slashed{\partial}\psi} \\ &\times\left(\prod_{I=1}^{N_++N_-}\int \frac{d^4z_IdU_I}{V}\Theta_I~\mathcal{O}_{\mathrm{QCD}}\right) 
    \end{aligned}
\end{equation}

The evaluation of the QCD operators from the instanton vacuum is two-fold. The quark degrees of freedom in the operators can be directly averaged by the effective Lagrangian in Eq.~\eqref{eq:effective_action}, while the gluon degrees of freedom are calculated by replacing the gauge field in the operator by the semi-classical background of the instantons \cite{Diakonov:1995qy,Weiss:2021kpt}. 

Thus, in the instanton ensemble, the averages over the  QCD operators convert the gluonic part into the corresponding effective quark degrees of freedom. The gluonic part of the operator distorts the color orientation of 't Hooft vertices, producing various quark spin structure, mapping the operators into numerous effective quark operators that can be evaluated further by the quark-instanton interactions in the instanton vacuum.

\begin{figure}
\centering
\subfloat[\label{fig:1g_1}]{\includegraphics[width=0.28\linewidth]{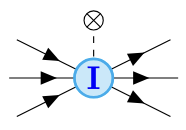}}
\hfill
\subfloat[\label{fig:2g_1}]{\includegraphics[width=0.29\linewidth]{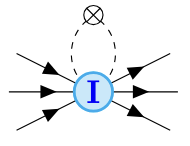}}
\hfill
\subfloat[\label{fig:2g_2}]{\includegraphics[width=0.4\linewidth]{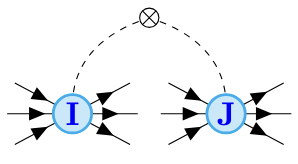}}
\hfill
\subfloat[\label{fig:3g_2}]{\includegraphics[width=\linewidth]{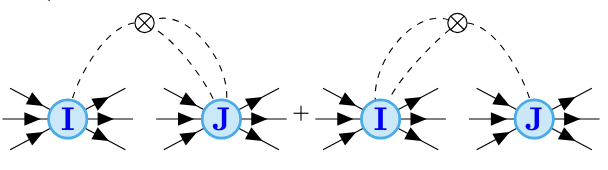}}
\caption{Feynman diagrams for QCD operators coupled to pseudoparticles. The crossdot denotes the operator $\mathcal{O}[A_{I_1},\cdots,A_{I_n}]$ in \eqref{eq:op_avg}. (a) one-gluon operators $\mathcal{O}[A_I]$ coupled to pseudoparticle $I$. (b) two-gluon operators $\mathcal{O}[A_{I_1},A_{I_2}]$ coupled to pseudoparticle $I$ twice or (c) coupled to two different pseudoparticle $I$ and $J$. (d) three-gluon operators $\mathcal{O}[A_{I_1},A_{I_2},A_{I_3}]$ coupled to two pseudoparticle $I$ with one pseudoparticle $J$ or one $I$ with two $J$.}
\label{fig:gluon_op}
\end{figure}

Using (\ref{eq:gluo_op}), the VEV of $\mathcal{O}_{\mathrm{QCD}}$ in the instanton ensemble can be organized 
in terms of the instanton density $n_{I+A}$.
\begin{widetext}
\begin{equation}
\label{eq:inst_contribution}
\begin{aligned}
    \langle \mathcal{O}_{\mathrm{QCD}}\rangle_{N_\pm}=&\sum_{n=1}^\infty\frac{1}{n!}\left[\sum_{k=0}^n\binom{n}{k}N_+^{n-k}N_-^{k}\langle\mathcal{O}_{++\cdots-}\rangle_{\mathrm{eff}}\right]\\[5pt]
    =&N_+\langle\mathcal{O}_+\rangle_{\mathrm{eff}}+N_-\langle\mathcal{O}_-\rangle_{\mathrm{eff}}+\frac{N^2_+}{2}\langle\mathcal{O}_{++}\rangle_{\mathrm{eff}}+N_+N_-\langle\mathcal{O}_{+-}\rangle_{\mathrm{eff}}+\frac{N^2_-}{2}\langle\mathcal{O}_{--}\rangle_{\mathrm{eff}}+\cdots
\end{aligned}
\end{equation}
where the effective fermionic operator $\mathcal{O}_{++\cdots-}$ is obtained by simultaneously connecting $\mathcal{O}_{\mathrm{QCD}}$ to the $n$ instantons by sharing the classical fields: 
\begin{equation}
\begin{aligned}
\label{eq:op_avg}
    \mathcal{O}_{++\cdots-}=&\left(\frac{1}{V( m^*)^{N_f}}\right)^n\underbrace{\int d^4z_{I_1}dU_{I_1}\cdots d^4z_{I_n}dU_{I_n}\mathcal{O}[\psi,\bar\psi,A_{I_1},A_{I_2},\cdots,A_{I_n}][\Theta_{I_1}\cdots\Theta_{I_n}]_{\rm conn}}_{ n-k~\mathrm{instantons}\,(I),~k~\mathrm{anti-instantons}\,(A)}
\end{aligned}
\end{equation}
\end{widetext}
where $\Theta_{I}$ is the 't Hooft vertices in \eqref{eq:tHooft}. 
This calculation can be graphically presented in Fig.~\ref{fig:gluon_op} with Feynman diagrams where the dashed lines trace the color exchange from pseudoparticles. The number of the dashed lines connected to the crossdot represents the number of gauge fields in the operator $\mathcal{O}[A_{I_1},\cdots,A_{I_n}]$. 

Given the complexity of the full calculation, the $1/N_c$ counting provides a useful way to organize terms and isolate the dominant contributions. Since each of the $UU^\dagger$ pair gives a $1/N_c$ factor in the large $N_c$ limit, each of the external quark (antiquark) lines and the dashed lines in the diagram contributes a pair of $UU^\dagger$ in the color group integral. The $1/N_c$ counting of each diagram is $1/N_c^{n_q+n_g-n_{\rm loop}}$ where $n_q$ is the open (unattached) quark number (the number of the open (anti)-quark lines in the diagram), $n_g$ is the gluon dashed lines, and the number of planar color flow loops $n_{\rm loop}$ formed by any quark or gluon lines passing through the vertices or cross dot. Those closed planar color loops contribute additional $N_c$ factors due to the color flow. 
With this in mind, the leading $1/N_c$ diagrams usually are the diagrams without external quark lines ($n_q=0$), corresponding to the disconnected diagrams in the matrix element \cite{Diakonov:1995qy}. When external hadron sources present, the disconnected diagrams still give nontrivial contributions due to the fluctuation in the instanton numbers. As a result, the canonical ensemble formulation has to be generalized to the grand canonical ensemble with varying $N_\pm$ (see Sec.~\ref{grand}).



Now the canonical ensemble average effectively reduces to the path integral of the effective quark field theory. The calculations become the VEVs of a bunch of effective quark operators over the effective Lagrangian in \eqref{eq:effective_action}. This is the consequences of the diluteness of the instanton vacuum. The calculations now can be done systematically order by order in the framework of the instanton density expansion. In this vacuum operator expansion, the series is naturally set around the energy scale $\rho^{-1}\simeq600$ MeV and controlled by the vacuum density $n_{I+A}$. The pseudoparticle profiles of each configurations (single-instantons and molecules) probed by the external momentum transfer $q$, which is zero for vacuum expectation but can be non-zero when hadron sources present (form factors). For more details, see \cite{Liu:2024rdm,Liu:2024jno,Liu:2024,Liu:2024yqa}


\subsection{Form factors}
\label{Sec:FF}
The arguments for vacuum averages can be extended to hadronic matrix elements, provided the resulting effective vertices remain localized within the hadronic scale. Given the comparable size of instantons and light hadrons, including the pion, the transition matrix element of the operator $\mathcal{O}_{\mathrm{QCD}}$ in a hadron state can be expressed as an ensemble average, similar to (\ref{eq:inst_contribution}), with vacuum bra-ket replaced by in-out on-shell hadronic states.
\bea
\label{eq:o_had_exp}
    \langle \mathcal{O}_{++\cdots-}\rangle_{N_\pm}\rightarrow \langle p'| \mathcal{O}_{++\cdots-}|p\rangle_{N_\pm} 
\eea

The form factors following from (\ref{eq:o_had_exp}) can be expanded systematically, in terms of the instanton density, which is commensurate with a  book-keeping in $1/N_c$. 
Translational symmetry  relates the hadronic matrix element of $\mathcal{O}_{\mathrm{QCD}}$ to the momentum transfer  between the hadronic states, 
\begin{equation}
     \langle p'|\mathcal{O}_{\mathrm{QCD}}|p\rangle = \frac{1}{V}\int d^4x\langle p'|\mathcal{O}_{\mathrm{QCD}}(x)|p\rangle e^{-iq\cdot x}
\end{equation}
The recoiling hadron momentum is defined as $p'=p+q$, and the forward limit follows from $q\rightarrow0$.
(\ref{eq:o_had_exp}) generalizes the arguments in~\cite{Weiss:2021kpt} to off-forward and multi-instanton
contributions. 

Graphically, the color-averaging in (\ref{eq:op_avg}) connects $\mathcal{O}_{\mathrm{QCD}}$ to $n$ instantons through the classical field backgrounds. Each matrix element in (\ref{eq:o_had_exp}) is evaluated by the effective Lagrangian in \eqref{eq:effective_action}, with only the connected diagrams retained. 
The calculations can be carried out order by order in the instanton density expansion due to the diluteness of the pseudoparticles in vacuum.


\section{Grand canonical instanton ensemble}
\label{grand}
Here the averaging over the fluctuations in the number 
of pseudoparticles in the ILM at zero theta vacuum angle is briefly outlined. In a
grand canonical description where $N_\pm$
are allowed to fluctuate with the measures
~\cite{Diakonov:1995qy,Schafer:1996wv,Zahed:2021fxk}
\begin{equation}
\label{DISTX}
  \mathds{\mathcal P}(N_+,N_-)=\mathbb P(N)\mathbb Q(\Delta) 
\end{equation}
with mean $\bar{N}=\langle N\rangle$ and  $Q_t=\langle\Delta\rangle=0$. 

The quantum scale fluctuations in QCD are captured in the instanton vacuum using the grand-canonical description, where the quasiparticle number $N=N_++N_-$ is allowed to fluctuate with the probability distribution
\cite{Diakonov:1995qy,Kacir:1996qn,Nowak:1996aj}
\be
\label{dist}
\mathbb P(N)=e^{\frac{bN}4 }\bigg(\frac {\bar{N}}{N}\bigg)^{\frac {bN}4 }
\ee
to reproduce the vacuum topological compressibility
\be
\label{COMP}
\frac{\sigma_t}{\bar{N}} =\frac{\langle(N-\bar N)^2\rangle}{\bar N}=\frac 4b \approx \frac 1{N_c}
\ee
in agreement with low-energy theorems~\cite{Novikov:1981xi}.
In this formulation the cooled QCD vacuum is a quantum liquid of pseudoparticles, with a topological compressibility of about $\frac 13$ at $N_c = 3$, and 
incompressible at large $N_c$.

The fluctuations in the number difference $\Delta$ are fixed by  the topological susceptibity~\cite{Diakonov:1995qy} 
\begin{equation}
\label{DISTQ}
    \mathds{Q}(\Delta)=
    \frac{1}{\sqrt{2\pi\chi_t}}\exp\left(-\frac{\Delta^2}{2\chi_t}\right)
\end{equation}
with the width of distribution defined by
\bea
\label{SUSQ}
\chi_t=\langle \Delta^2\rangle
\eea

This quenched topological susceptibility can be related to the meson singlet mass by the Witten-Veneziano formula~\cite{Witten:1979vv,Veneziano:1979ec}
\be
\label{CHI}
\frac{\chi_t}V=\frac{f_\pi^2}{2N_f}\left(m_{\eta^\prime}^2+m_\eta^2-2m_K^2\right)
\ee
This implies that the mass of $\eta'$ is essentially due to the axial anomaly relating to non-trivial topological charge fluctuations, which can turn out to be nonzero even in the chiral limit.

A naive estimation in quenched vacuum predicted by \cite{Diakonov:1983hh} with simple sum ansatz is
\bea
\label{SUSQ}
\chi_t= \bar N
\eea
However, this result contradicts large-$N_c$ scaling that $\chi_t\sim \mathcal{O}(N_c^0)$ whereas $\bar{N}\sim \mathcal{O}(N_c)$, indicating a flaw in sum ansatz \cite{hutter2001instantons}. In general, the quenched topological susceptibility is subject to interaction between instantons and anti-instantons. A general derivation with Feynman variation principle in \cite{Diakonov:1995qy} gives

\begin{equation}
\label{chi_t}
\chi_t= \frac{4\bar N}{b-\frac{\gamma_a}{\gamma_s}(b-4)}
\end{equation}
where $b=11N_c/3$ and $\gamma_{s,a}$ is the mean interaction between pseudoparticle pairs with the same duality and the ones with opposite duality.

\begin{equation}
    \int d^4Rdu\, S_{int}(R,u,\rho_1,\rho_2)=\frac{8\pi^2}{g^2}\rho_1^2\rho_2^2\gamma_{s,a}
\end{equation}
where $\gamma_{s,a}	\propto N_c/(N_c^2-1)$. For $\gamma_a=0$, the result in \eqref{chi_t} reduces to the result obtained from exact classical solutions, indicating the nontrivial interaction between $IA$ pairs break the duality. It also indicates that an attractive interaction decreases the susceptibility, whereas a repulsive interaction would increases it.

\begin{eqnarray}
\chi_t&<\;\dfrac{4}{b}\,\bar{N},\, & \text{for attractive } IA\nonumber\\[6pt]
\chi_t&>\;\dfrac{4}{b}\,\bar{N},\, & \text{for repulsive } IA
\end{eqnarray}

A more sophisticated method using streamline ansatz \cite{1991NuPhB.362...33V} has shown that the strong $IA$ repulsion is an artifact of the simple sum ansatz. The $IA$ interaction strongly depends on the
orientation and the average repulsion is about 14 times smaller than the one obtained in \cite{Diakonov:1983hh}. Therefore, one can practically assume the average $IA$ interaction to be small such that quenched susceptibility can be approximated by

\begin{equation}
\label{ILM_chi}
    \chi_t\approx\frac4b \bar{N}
\end{equation}

In (unquenched) QCD, light quarks introduce strong attraction between instantons and anti-instantons. As a result, $\chi_t$ is substantially screened by the light quarks~\cite{Diakonov:1995qy,Kacir:1996qn},
\bea
\label{SUS}
\frac{\chi_t}{\bar N}\sim 
\bigg(\frac{11}{12}N_c-\frac{n_{I+A}}{\langle\bar qq\rangle}\sum_f\frac{1}{m_f}\bigg)^{-1}
\eea
This result agrees with the chiral perturbation calculation in chiral limit \cite{Leutwyler:1992yt}. With the identification of quark condensate $\langle \bar q q\rangle=-n_{I+A}/m^*$, the result in \eqref{SUS} can be further simplified to

\begin{equation}
\frac{\chi_t}{\bar N}\sim 
\bigg(\frac{11}{12}N_c+N_f\frac{m^*}{m}\bigg)^{-1}
\end{equation}

\begin{figure}
    \centering
    \includegraphics[width=\linewidth]{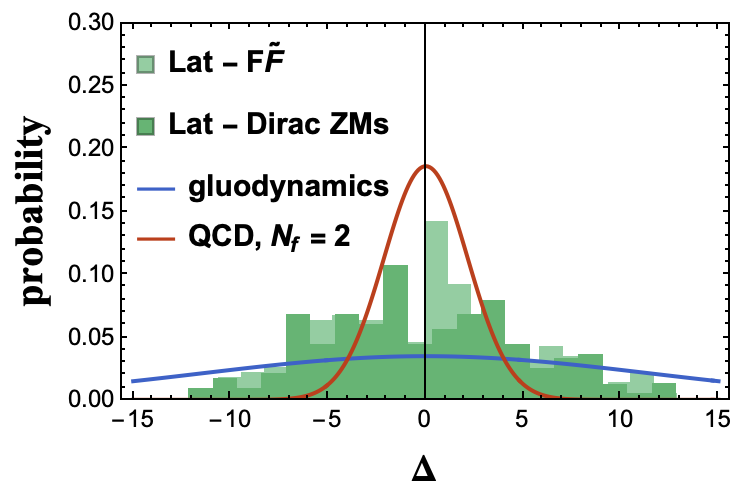}
    \caption{The ILM result is compared with $\chi$QCD lattice calculation using overlap fermions with the lattice size: $24^3\times64 ~a^4$ where $a=0.1105$ fm \cite{Liang:2023jfj,Bhattacharya:2021lol,Alexandrou:2020mds}. The distribution presented by the light green obtained by propagating the gluonic operator $F\tilde{F}$ with large lattice flow time $t_f =4a^2$ and the distribution presented by the dark green calculates the topological charges from counting the Dirac zero modes on the lattice.}
    \label{fig:topo}
\end{figure}

\begin{figure}
    \centering
    \includegraphics[width=\linewidth]{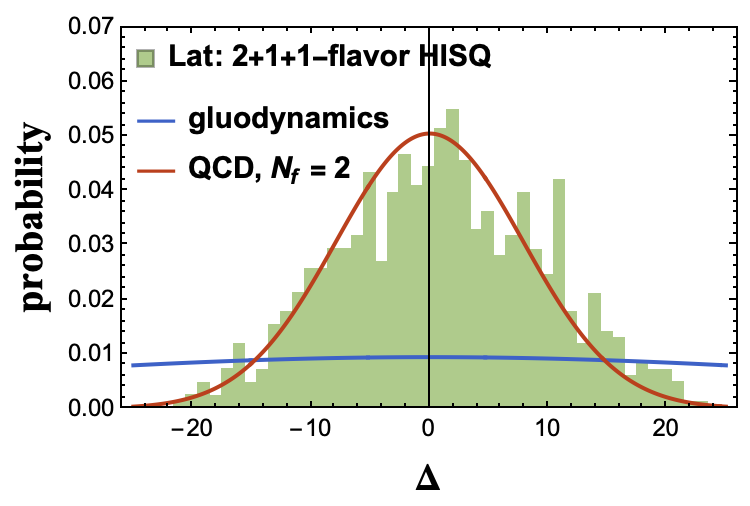}
    \caption{The ILM result is compared with lattice calculation using HISQ ensemble with physical pion mass $m_\pi=135$ MeV and lattice volume: $96^3\times192 ~a^4$ where the lattice spacing is $a=0.0570$ fm \cite{Bhattacharya:2021lol}.}
    \label{fig:topo-2}
\end{figure}

\begin{figure}
    \centering
    \includegraphics[width=\linewidth]{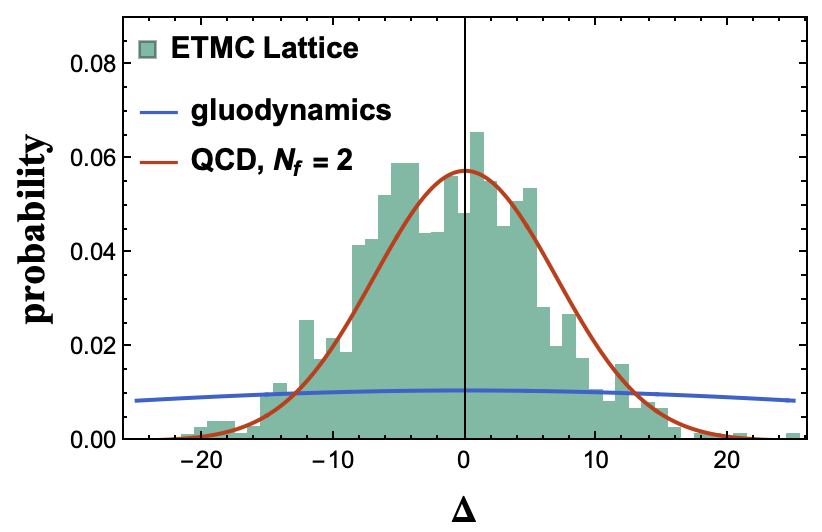}
    \caption{The ILM result is compared with ETMC lattice calculation using $N_f = 2+ 1+ 1$ twisted mass clover-improved fermions with the lattice size: $64^3 \times 128~a^4$ where \textcolor{black}{$a=0.0801(4)$ fm} and physical pion mass $m_\pi=139$ MeV \cite{Alexandrou:2020mds}. }
    \label{fig:topo-3}
 \end{figure}

With topological screening, the Witten-Veneziano formula in \eqref{CHI} is also modified by the substitution of the singlet mass with pion mass $m_\pi^2$. Table \ref{tab:chi} presents the estimation of susceptibility with Witten-Veneziano formula and ILM in both quenched vacuum and $N_f=2+1$ QCD vacuum.
\begin{table}[]
    \centering
    \begin{tabular}{c|c|c|c}
        \hline
        $\chi_t$ & WV \eqref{CHI} & ILM \eqref{SUSQ} & ILM \eqref{ILM_chi} \\
        \hline
       quenched &$(180.2\,\mathrm{MeV})^4$ & $(197.3\,\mathrm{MeV})^4$ & $(153.2\,\mathrm{MeV})^4$ \\
       
      $N_f=2+1$ & $(72.6\,\mathrm{MeV})^4$ & $(63.9\,\mathrm{MeV})^4$ & $(63.6\,\mathrm{MeV})^4$ \\
        \hline
    \end{tabular}
    \caption{WV denotes the result from Witten-Veneziano formula in \eqref{CHI} using $m_\pi=139$ MeV,$m_K=494$ MeV, $m_{\eta}=549$, and $m_{\eta'}=958$ MeV from Particle Data Group \cite{ParticleDataGroup:2024cfk}. ILM denotes the instanton estimation from \eqref{SUSQ} and \eqref{ILM_chi} using $n_{I+A}=1$ fm$^{-4}$, $m_{u}\langle\bar{u}u\rangle=m_{d}\langle\bar{d}d\rangle=-3.4\times10^{-5}$ \cite{FlavourLatticeAveragingGroupFLAG:2021npn} and $m_{s}\langle\bar{s}s\rangle=-1.7\times10^{-3}$ \cite{Harnett:2021zug}.  }
    \label{tab:chi}
\end{table}
In Fig.~\ref{fig:topo}, \ref{fig:topo-2}, and \ref{fig:topo-3}, the prediction in this work is compared with instanton liquid ensemble and the lattice calculation performed by three different groups. The blue curve denotes the quenched calculation in instanton liquid ensemble with $\chi_t/V=(197~\mathrm{MeV})^4$ and the red curve denotes the 2-flavor QCD instanton liquid ensemble where the susceptibility is manually set to be $\chi_t/V=(85~\mathrm{MeV})^4$. Various lattice results \cite{Liang:2023jfj} indicate the Gaussian form the topological charge distribution and are consistent with ILM prediction.

\subsection{Theta vacuum}

In the QCD vacuum, topologically active pseudoparticles, namely instantons and anti-instantons, are CP-conjugated pairs, which naturally serve as sources of local CP violation. At a finite $\theta$-angle, CP symmetry is significantly violated in QCD. Using the ensemble measure provided in \eqref{DISTX}, the grand canonical partition function at a finite vacuum angle can be reconstructed with parameterization in terms of $\mu$ and $\theta$.

\begin{equation}
\begin{aligned}
    \mathcal{Z}(\mu,\theta)=&\sum_{N,\Delta}Z_{N_\pm}e^{\mu N+i\theta \Delta}\\
    =&\exp\left(\frac b4\langle N\rangle_\theta e^{\frac{4\mu}{b}}\right)
\end{aligned}
\end{equation}
The chemical potential $\mu$ eventually will be tuned to zero as instantons and anti-instantons correspond to zero energy tunneling. Meanwhile, the $\theta$ term acts as a topological chemical potential for the topological charge $\Delta$, enhancing instantons and depleting anti-instantons. The angle dependence of $\langle N\rangle_\theta$ in general is subject to the interaction between pseudoparticles and the presence of light quarks. For simple sum ansatz $\gamma_a=\gamma_s$ in quenched vacuum with finite theta, the angle dependence is determined by


\begin{align}
\label{SCALE}
\langle N\rangle_\theta=\bar{N}(\cos\theta)^{4/b}
\end{align}

For $\gamma_a=0$, the angle dependence instead becomes

\begin{align}
\label{SCALE}
\langle N\rangle_\theta=\bar{N}\cos\left(\frac4b\theta\right)
\end{align}

The general solution for quenched vacuum obtained by Feynman variational principle is

\[
\langle N\rangle_\theta
=
\cos\vartheta
\left(
\frac{
\gamma_a\cos\vartheta + \sqrt{\gamma^2_s - \gamma_a^2 \sin^2\vartheta}
}{
\gamma_s+\gamma_a
}
\right)^{\frac{4-b}{b}}
\]
where the angle $\vartheta$ can be controlled by
\[
\vartheta + \frac{b-4}{4}\,
\arctan\!\left(
\frac{(\gamma_s^2-\gamma_a^2)\sin\vartheta}{\gamma_s^2\cos\vartheta + \gamma_a\sqrt{\gamma_s^2-\gamma_a^2\sin^2\vartheta}}
\right)
= \theta
\]

For more details, see \cite{Diakonov:1995qy}.

\subsection{Vacuum expectation value}

As a result of the instanton liquid ensemble, most QCD operators are averaged through
\begin{equation}
\begin{aligned}
\label{grand_canonical}
    \langle \mathcal{O}_{\mathrm{QCD}}\rangle=&\sum_{N_+,N_-}\mathcal {P}(N_+,N_-)\langle \mathcal{O}_{\mathrm{QCD}}\rangle_{N_\pm}
    \\
    \equiv& \overline{\langle\mathcal{O}_{\mathrm{QCD}}\rangle}_{N_\pm} 
\end{aligned}
\end{equation}
The  averaging is carried  over the configurations with fixed $N_\pm$ (canonical ensemble average),  followed by an averaging over the distribution (\ref{DISTX}).

A well-known calculation within the ILM framework is the determination of the gluon condensate. In the canonical ensemble, the vacuum average of the gluonic scalar operator is directly proportional to the total number of instantons, $N$. 
\begin{equation}
    \frac1{32\pi^2}\langle F^2_{\mu\nu}\rangle_{N_\pm}= \frac NV
\end{equation}

Thus, after taking the fluctuations into account, the VEV of gluonic scalar operator, namely gluon condensate, corresponds to the instanton density $n_{I+A}$.


The same calculation applies to the gluonic pseudoscalar operator. Its vacuum average in canonical ensemble is proportional to the total instanton number difference $\Delta$.

\begin{equation}
\frac1{32\pi^2}\langle F_{\mu\nu}\tilde{F}_{\mu\nu}\rangle_{N_\pm}= \frac \Delta V
\end{equation}

However, after taking the fluctuations into account, the VEV of gluonic pseudoscalar operator becomes zero, indicating that the QCD vacuum is topologically neutral.


\subsection{Hadronic matrix element}

The similar calculation can be applied to the evaluation of the hadronic matrix elements. Yet the calculation is more involved. It can be formally written as a large time reduction of a  3-point function
\begin{equation}
    \frac{\langle h|\mathcal{O}_{\mathrm{QCD}}|h\rangle}{\langle h|h\rangle}=\lim_{t\rightarrow\infty}\frac{\langle J^\dagger_h(t/2)\mathcal{O}_{\mathrm{QCD}} J_h(-t/2)\rangle}{\langle J^\dagger_h(t/2)J_h(-t/2)\rangle}
\end{equation}
where $J_h(t)$ is a pertinent source for  the hadronic state $h$ defined by

\begin{equation}
    J_h(p,t)=\int d^3\vec{x} e^{-i\vec{p}\cdot \vec{x}}J_h(x)
\end{equation}


In the setting of the canonical ensemble with \eqref{eq:op_avg}, the diagrams leading in $1/N_c$ counting are usually the diagrams disconnected to the hadron sources, resulting in no contribution to the hadron matrix element.

\begin{equation}
\begin{aligned}
    &\langle J^\dagger_h(t/2)\mathcal{O}_{\rm QCD}J_h(-t/2)\rangle_{N_\pm}\\
    =&\langle J^\dagger_h(t/2)J_h(-t/2)\rangle_{N_\pm}\langle\mathcal{O}_{\rm QCD}\rangle_{N_\pm}\left(1+\mathcal{O}(1/N_c)\right)
\end{aligned}
\end{equation}

However, these diagrams are subject to the vacuum fluctuation of the topological pseudoparticles, which is not included in canonical ensemble. By extending the calculation to the grand canonical framework, the ensuing 3-point correlation function is carried out by


\begin{widetext}
\begin{equation}
\begin{aligned}
\langle J^\dagger_h(t/2)\mathcal{O}_{\mathrm{QCD}} J_h(-t/2)\rangle
=&\sum_{N_+,N_-}\mathcal{P}(N_+,N_-)\left[\langle\mathcal{O}_{\rm QCD}\rangle_{N_\pm}-\overline{\langle\mathcal{O}_{\rm QCD}\rangle}_{N_\pm}\right]\langle J_h^\dagger(t/2) J_h(-t/2)\rangle_{N_\pm}
\end{aligned}    
\end{equation}
By expanding the fluctuation to the leading order and implementing the asymptotic Euclidean time limit,
$$\lim_{t\rightarrow\infty}\left\langle J^\dagger_h(t/2)J_h(-t/2)\right\rangle_{N_\pm}\rightarrow e^{-m_h(N_+,N_-)t}$$
the hadronic matrix element reads
\begin{equation}
\begin{aligned}
    \frac{\langle h|\mathcal{O}|h\rangle}{V}
    =&-\overline{\left[\langle\mathcal{O}_{\rm QCD}\rangle_{N_\pm}-\overline{\langle\mathcal{O}_{\rm QCD}\rangle}_{N_\pm}\right](N-\bar{N})}\left(\frac{\partial m^2_h}{\partial N}\right)\bigg|_{\substack{N=\bar{N}\\ \Delta =0}}\\
    &-\overline{\left[\langle\mathcal{O}_{\rm QCD}\rangle_{N_\pm}-\overline{\langle\mathcal{O}_{\rm QCD}\rangle_{N_\pm}}\right]\Delta}\left(\frac{\partial m^2_h}{\partial \Delta}\right)\bigg|_{\substack{N=\bar{N}\\ \Delta =0}}
\end{aligned}
\end{equation}
\end{widetext}

One can directly apply this calculation to the matrix element of gluonic scalar and pseudoscalar operators at the leading $1/N_c$. The scalar gluon matrix elemenet is tied to the topological compressibility,
\begin{equation}
\begin{aligned}
    \frac1{32\pi^2}\langle h|F^2_{\mu\nu}|h\rangle
    =-2m^2_h \sigma_t
    \frac{\partial\ln m_h}{\partial N}\bigg|_{\substack{N=\bar{N}\\ \Delta =0}}
\end{aligned}
\end{equation}
and the matrix element of pseudoscalar gluon at the leading $1/N_c$ is tied to the topological susceptibility. 
\begin{equation}
\begin{aligned}
    \frac1{32\pi^2}\langle h|F_{\mu\nu}\tilde{F}_{\mu\nu}|h\rangle
    =-2m^2_h \chi_t
    \frac{\partial\ln m_h}{\partial \Delta}\bigg|_{\substack{N=\bar{N}\\ \Delta =0}}
\end{aligned}
\end{equation}

This general framework can provide a robust framework for vast applications in calculations of various hadronic matrix element \cite{Liu:2024rdm,Liu:2024jno,Liu:2024sqj,Liu:2024vkj,Weiss:2021kpt,Diakonov:1995qy,Kim:2023pll}.


\section{Form factors in different probing scale $Q^2$}
\label{FF}
In general, the form factor can be described by the hadronic states with the probe defined by QCD operators $\mathcal{O}_{\rm QCD}$. To calculate the form factor, one need the information of hadron states and the probe. 

\begin{figure}
    \centering
    \includegraphics[width=1\linewidth]{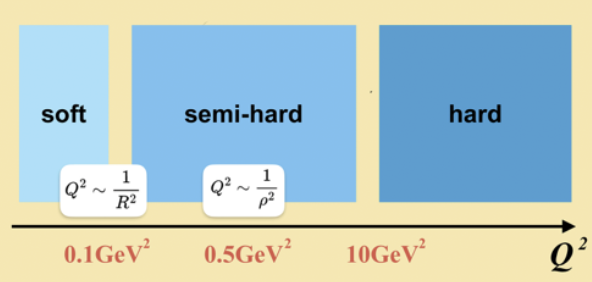}
    \caption{The soft, semi-hard, and hard $Q^2$ energy regions characterizing different regimes of the hadronic form factors}
    \label{fig:scales}
\end{figure}

As illustrated in Fig.~\ref{fig:scales}, the momentum scale $Q^2$ probing the hadronic structure is approximately divided into three regions: soft, semi-hard, and hard region. 

\subsection{Hard region}

At large momentum transfer $Q\rho\gg 1$, or the hard region, the meson form factors are fixed by factorization and the perturbative hard scattering. The factorization can be naturally described by Breit (brick-wall) frame,

\begin{align*}
    p_\mu=&(Q/\sqrt{2},0^-,0_\perp) &
    p'_\mu=&(0^+,Q/\sqrt{2},0_\perp)
\end{align*}
where momentum $p$ recoils as $p'$ in the opposite direction under the probe $Q$. This particular momentum configuration presents the fast moving hadron in large $Q^2$, allowing the factorization to be formulated in terms of perturbative hard kernels, $ T(x,x',\mu/Q, Q^2)$, and light cone observables, namely the nonperturbative hadron distribution amplitudes, $\varphi(x, \mu) $ (longitudinal momentum light-cone wave functions), at a specified renormalization scale $\mu $, where $ Q \gtrsim \mu \gg \Lambda_{\rm QCD}$.

\begin{figure}
    \centering
\subfloat[\label{fig:m_FF}]{\includegraphics[width=1\linewidth]{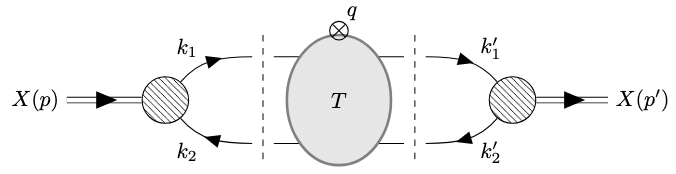}}
\hfill
\subfloat[\label{fig:b_FF}]{\includegraphics[width=1\linewidth]{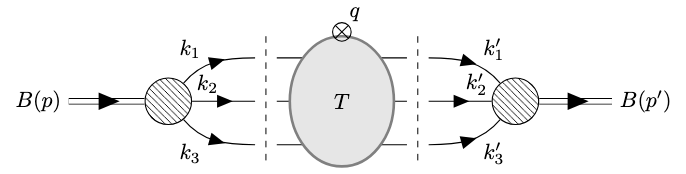}}
    \caption{The factorization of (a) meson and (b) baryon form factors at large $Q^2$}
\label{fig:FF}
\end{figure}

As illustrated in Fig.~\ref{fig:FF}, the hadron form factor reads  \cite{Braaten:1987yy,Sterman:1997sx,Dittes:1983dy},

\begin{equation}
\begin{aligned}
\label{eq:LF_fac}
&\langle p' |\mathcal{O}_{\rm QCD}| p\rangle=\\
&\int dx\int dx'\varphi^*(x,\mu) T(x,x',\mu/Q,Q^2)\varphi(x',\mu)
\end{aligned}
\end{equation}
where the integration over the longitudinal momentum fraction is defined as $$dx=dx_1dx_2\delta(1-x_1-x_2)$$ for meson and $$dx=dx_1dx_2dx_3\delta(1-x_1-x_2-x_3)$$ for baryon. 

\begin{figure}
    \centering
\subfloat[\label{T1}]{\includegraphics[width=0.5\linewidth]{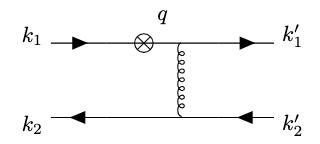}}
\hfill
\subfloat[\label{T3}]{\includegraphics[width=0.5\linewidth]{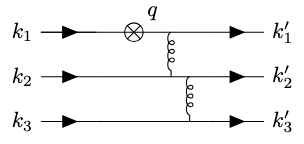}} 
\hfill
\subfloat[\label{T2}]{\includegraphics[width=0.5\linewidth]{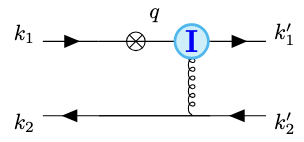}} 
\hfill
\subfloat[\label{T4}]{\includegraphics[width=0.5\linewidth]{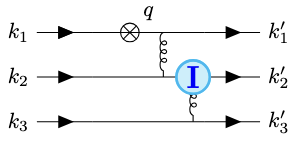}} 
    \caption{(a) and (b) are the perturbative gluon exchange diagrams at the leading order of $\alpha_s(Q^2)$. (c) and (d) are the small instanton-induced chromomagnetic gluon exchange in the hard kernel $T$.}
    \label{fig:T}
\end{figure}

For the hard kernel $T$, besides from the perturbative gluon exchange, at intermediate $Q^2$, the small sized instantons  $\rho\sim1/Q$ can also contribute to the hard scattering  \cite{Kochelev:1996pv,Dorokhov:2003dt,Korchagin:2011wy,Ringwald:1998wp}. The induced anomalous chromomagetic gluon exchange flips the quark chirality, presenting a novel vertex distinct from perturbative gluon exchange, of \textcolor{black}{which} \textcolor{black}{the strength $4\pi\kappa_{I+A}/N_c$ is  comparable to the 1-loop perturbative result in $\alpha_s(Q^2)$} \cite{Dorokhov:2004fb,Dorokhov:2002qf}. In Fig.~\ref{fig:T}, the hard kernel diagrams with perturbative gluon exchange at the leading order of $\alpha_s$ and instanton-induced gluon exchange is shown.

Incorporating twist expansion of the distribution amplitudes, this factorization formulation can be systematically extended to include power corrections suppressed by $1/Q^2$, thereby enhancing the accuracy by including subleading power corrected contributions~\cite{Shuryak:2020ktq,Liu:2024vkj}. Higher-order perturbative corrections often results in a logarithmic scale dependence of the form $\ln^n(\mu^2/Q^2)$ in the hard kernel, arising from the truncation at finite order. To avoid large logarithms from \textcolor{black}{compromising} the validity of factorization, a renormalization scheme such that $\mu\sim Q$ is often assumed.

At asymptotic limit ($Q^2\rightarrow\infty$), the form factor follows $n-$pole form by Brodsky-Farrar (constituent quark) counting rule \cite{Brodsky:1973kr} where $n$ is the minimum number of the parton spectators. In the case of electromagnetic form factor, it is monopole for meson and dipole for baryon.
\begin{equation}
\langle p' |\mathcal{O}_{\rm QCD}| p\rangle\sim\frac1{(Q^2)^n}
\end{equation}

\begin{figure}
    \centering
\subfloat[\label{fig:WF_m}]{\includegraphics[width=\linewidth]{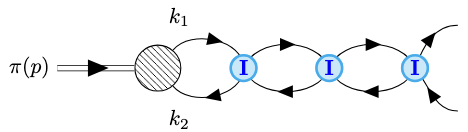}}
\hfill 
\subfloat[\label{fig:WF_r}]{\includegraphics[width=\linewidth]{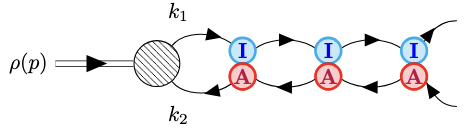}}
\hfill 
\subfloat[\label{fig:WF_b}]{\includegraphics[width=\linewidth]{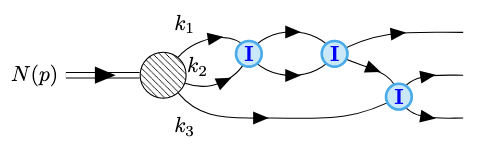}}
    \caption{The hadron portrayed at renormalization scale $\mu\lesssim1/\rho$ passing through the QCD instanton vacuum for (a) pion, (b) rho meson, (c) nucleon. The gray shaded blob represents the hadronic BS amplitude $\Gamma_h$ obtained by amputating the external constituent quark in $\Psi_h$ 
    (See text)}
    \label{fig:Had}
\end{figure}

\subsection{Semi-hard and soft region}

At small momentum transfer $Q^2$, the factorization begins to lose its validity. The presence of pseudoparticles alters the point-like probe vertex.  \textcolor{black}{As a result, the factorization picture of a hadron at large $Q^2$} transitions into a constituent quark representation at low $Q^2$.
\textcolor{black}{In contrast to the large-\(Q^2\) light-front factorization framework formulated in Minkowski space, in this section we adopt a Euclidean construction that is naturally suited to the instanton-liquid description. An equivalent light-front formulation consistent with~\eqref{eq:LF_fac} can be obtained through an appropriate analytical continuation. For more details, see~\cite{Liu:2023yuj,Liu:2023fpj}.}

\textcolor{black}{
In this regime, most perturbative gluons are depleted, leaving behind a highly heterogeneous vacuum in which constituent quarks propagate, providing a pictorial description where the hadron dynamics are primarily determined by the emerging 't Hooft interaction \eqref{eq:tHooft} between constituent quarks and pseudoparticles when passing through the vacuum as illustrated in Fig.~\ref{fig:Had}.}

These interactions can be systematically organized by resumming the leading $1/N_c$ $s$-channel bubble diagrams among the constituent quarks in the \(1/N_c\) book-keeping, giving rise to the hadronic Bethe-Salpeter (BS) wave function \( \Psi_h(k; p) \), given by

\begin{equation}
\begin{aligned}
\Psi_h(k_1\cdots k_n;p)
&=
S(k_1)\,\cdots
S(k_n)\,
\Gamma_{h}
\end{aligned}
\end{equation}
\textcolor{black}{where $\Gamma_h$ is the hadronic BS amplitude  obtained by amputating the external constituent quarks.}



\begin{figure*}
    \centering
\subfloat[\label{fig:hard_m}]{\includegraphics[width=.49\linewidth]{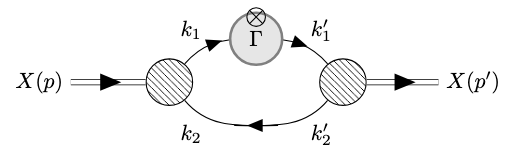}}
\hfill
\subfloat[\label{fig:hard_b}]{\includegraphics[width=.49\linewidth]{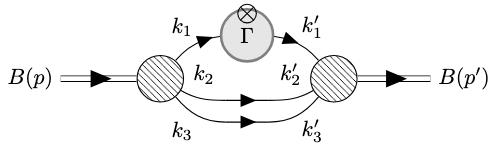}}
    \caption{The form factors of (a) meson $X$ and (b) baryon $B$ at small $Q^2$ with hadron state addressed by the BS wave function. The probe denoted by the cross dot is dressed by the instanton vacuum (See text). The momentum conservation requires $k_1'=k_1+q$ and $k_{2,3}'=k_{2,3}$. }
    \label{fig:WF_FF}
\end{figure*}

With this wave function alongside the effective probe at different low $Q^2$ regime, the small $Q^2$ form factor for the meson, as illustrated in Fig.~\ref{fig:hard_m} can be computed by  

\begin{widetext}

\begin{equation}
\begin{aligned}
\langle X(p') |\bar{\psi}\Gamma \psi| X(p)\rangle=&\int d[12]\mathrm{Tr}\bigg[\overline{\Psi}_X(k_1+q,k_2;p')\Gamma(k_1+q,k_1)\Psi_X(k_1,k_2;p)S^{-1}(-k_2)\bigg]\\
&+\int d[12]\mathrm{Tr}\bigg[\Psi_X(k_1,k_2;p)\Gamma(-k_2,-k_2-q) \overline{\Psi}_X(k_1,k_2+q;p')S^{-1}(-k_1)\bigg]
\end{aligned}
\end{equation}
and for baryons as illustrated in Fig.~\ref{fig:hard_b}, 
\begin{equation}
\begin{aligned}
\langle B(p') |\bar{\psi}\Gamma \psi| B(p)\rangle=&\int d[123]\mathrm{Tr}\bigg[\overline\Psi_B(k_1+q,k_2,k_3;p')\Gamma(k_1+q,k_1) S^{-1}(k_2)S^{-1}(k_3)\Psi_B(k_1,k_2,k_3;p)\bigg]\\
&+\int d[123]\mathrm{Tr}\bigg[\overline\Psi_B(k_1,k_2+q,k_3;p')\Gamma(k_2+q,k_2) S^{-1}(k_1)S^{-1}(k_3)\Psi_B(k_1,k_2,k_3;p)\bigg]\\
&+\int d[123]\mathrm{Tr}\bigg[\overline\Psi_B(k_1,k_2,k_3+q;p')\Gamma(k_3+q,k_3) S^{-1}(k_1)S^{-1}(k_2)\Psi_B(k_1,k_2,k_3;p)\bigg]
\end{aligned}
\end{equation}
\end{widetext}
where $\overline\Psi_{X,B}$ is the conjugate wave \textcolor{black}{function} for meson $X$ or baryon $B$ and $n$-body loop-momentum integral is defined as
$$d[1\cdots n]=\prod_{i=1}^n\int \frac{d^4k_i}{(2\pi)^4}(2\pi)^4\delta^4(p-\sum_{i}^nk_i)$$ 
\textcolor{black}{with the momentum integrals are eventually Wick-rotated to Euclidean space for the further evaluation.}

\textcolor{black}{In this regime, one can distinguish between semi-hard and soft domain using intrinsic ILM scale $\mu_0\sim1/\rho$. For the semi-hard domain, characterized by $Q\sim \mu_0$, single instantons and correlated instanton–anti-instanton pairs can be resolved by the external probe~\cite{Shuryak:2020ktq}. On the other hand, in the soft domain \(Q\ll \mu_0\), the probe is insensitive to individual pseudoparticles, and the underlying  collective instanton dynamics is more naturally described in terms of meson and glueball exchanges.} 


\subsubsection{Probe in semi-hard region}

\begin{figure}
    \centering
\subfloat[\label{fig:v_1}]{\includegraphics[width=.5\linewidth]{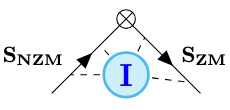}}
\hfill
\subfloat[\label{fig:v_2}]
{\includegraphics[width=.5\linewidth]{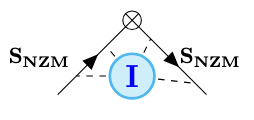}}
    \caption{(a) Quark probe $\Gamma$ dressed by the mix of zero and nonzero quark modes, resulting in a chirality-flipping vertex. (b) Quark probe $\Gamma$ dressed by nonzero quark mode alone, resulting in a chiral-preserving vertex.}
    \label{fig:semi-hard}
\end{figure}

In {\bf semi-hard} regime \textcolor{black}{($Q\sim \mu_0\sim 1/\rho$)}, single instantons and instanton pairs modify the probe vertex as shown in Fig.~\ref{fig:semi-hard}. Mostly the zero mode contributions are dominant in this regime and are fully taken care by vacuum operator expansion in \eqref{eq:inst_contribution} with the effective Lagrangian \eqref{eq:tHooft_g}. 

However, in certain situations, the contribution from nonzero modes is non-negligible. As shown in Fig.~\ref{fig:v_1}, the nonzero-mode contribution alone gives rise to a chirality-preserving effective vertex, whereas its interference with zero modes leads to the emergence of a chirality-flipping effective vertex.



Now with the single instanton effect, the quark probe vertex $\Gamma$ modified by the pseudoparticles reads

\begin{widetext}
\begin{equation}
\begin{aligned}
\label{VIIBAR}
   \Gamma_\pm(x,x')=\int dz_I dU_I\int d^4ye^{-iq\cdot y} &i\slashed{\partial}_x\bigg[\left(S_{\mathrm{NZM}}(x-z_I,y-z_I)-S_0(x-y)\right)\Gamma S_{\mathrm{ZM}}(y-z_I,x'-z_I)\\
   &+ S_{\mathrm{ZM}}(x-z_I,y-z_I)\Gamma \left(S_{\mathrm{NZM}}(y-z_I,x')-S_0(y-x')\right)\\
   &+ S_{\mathrm{NZM}}(x-z_I,y-z_I)\Gamma S_{\mathrm{NZM}}(y-z_I,x'-z_I)\bigg]i\slashed{\partial}_{x'}
\end{aligned}
\end{equation}
\end{widetext}
where the non-zero mode propagator $S_{\rm NZM}$ is defined in \eqref{eq:NZM}. The zero mode propagator $S_{\rm ZM}$, on the other hand, is defined
\begin{equation}
\label{ZMODE}
    S_{\mathrm{ZM}}(x,y)=\frac{\phi_I(x)\phi_I^\dagger(y)}{im}
    \rightarrow\frac{\phi_I(x)\phi_I^\dagger(y)}{im^*}
\end{equation}
The singular $1/m$ in the single instanton zero modes is shifted to finite $1/m^*$ by disordering in the multi-instanton background~\cite{Pobylitsa1989TheQP,Schafer:1996wv} (See also Sec.~\ref{QP} and references therein).

The on-shell reduction scheme of the in-out quark lines can be simplified by the zero momentum approximation as detailed in appendix~\ref{App:reduction_large_time}.
With this in mind, the modified probe vertex is 
\begin{equation}
  \Gamma \rightarrow\frac{N_+}{V}\Gamma_+(k',k)+\frac{N_-}{V}\Gamma_-(k',k)
\end{equation}

The point-like probe vertex becomes non-local distorted by the quark zero modes and non-zero modes.

\subsubsection{Probe in soft region}

\begin{figure}
    \centering
    \subfloat[\label{fig:soft_2}]{%
    \includegraphics[width=.5\linewidth]{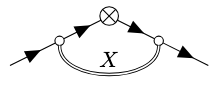}%
}\hfill
    \subfloat[\label{fig:soft_4}]{%
    \includegraphics[width=.32\linewidth]{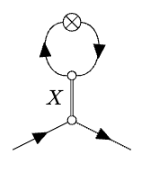}%
}
    \caption{Soft contributions 
to the hadron form factors where the instanton-induced $t$-channel bubble resummation forms streaks around the probe: (a) one-loop meson cloud dressing and (b) meson $X$ $t$-channel exchange}
    \label{fig:soft_FF}
\end{figure}

In {\bf soft} region \textcolor{black}{($Q\ll \mu_0\sim 1/\rho$)}, the single instanton approximation becomes invalid as the probe interact simultaneously with multiple pseudoparticles. 
As a result, meson and glueball exchanges induced by the emerging instanton interactions are dominant~\cite{Liu:2024rdm,Liu:2023fpj}, modifying the probe as illustrated in Fig.~\ref{fig:soft_FF}. 

The resummation of collective interactions among pseudoparticles gives rise to effective glueball-quark or meson-quark interaction given in \eqref{semi-bos}, and often result in a "pion cloud" surrounding most hadrons. In this regime, the form factors are mostly described by phenomenological chiral Lagrangians and gluodynamics, characterized by hadronic parameters such as mass spectrum or hadronic couplings.

For more details and complete results on different form factors compared to experimental measures and lattice calculations, see \cite{Liu:2024jno,Liu:2024sqj,Liu:2024vkj,Liu:2024rdm} (and references therein).

\section{Conclusion}

\textcolor{black}{This work has reviewed} the vacuum topological structures such as instantons, dyons, and center vortices. Their RG evolution across different energy scales are studied through several techniques on lattice QCD, revealing extraordinary topological landscape in the QCD vacuum. Among these structures, center vortices are believed to play a crucial role in the confinement mechanism and instantons serve as the underlying drivers of chiral and conformal symmetry breaking. By establishing a \textcolor{black}{instanton liquid} framework based on the topological vacuum structure, a systematic non-perturbative approach to the calculation of VEVs, hadron matrix elements, and form factors has been provided, thereby offering a robust theory for analyzing QCD phenomena \textcolor{black}{nonperturbatively}. Furthermore, our extension of the canonical instanton liquid ensemble to the grand canonical framework enables a more comprehensive treatment of topological vacuum fluctuations, enhancing the predictive power of the framework. \textcolor{black}{By reviewing recent progress across this series of works, a foundation has been established for a deeper understanding of the QCD vacuum structure and its fundamental role in the emergence of hadronic properties.}

\begin{acknowledgments}\noindent
The work of WYL is supported by the U.S. Department of Energy, Office of Science, Office of Nuclear Physics under Contract No. DE-FG-88ER40388. This work is also supported in part by the Quark-Gluon Tomography (QGT) Topical Collaboration, with Award DE-SC0023646.
\end{acknowledgments}

\appendix

\section{Conventions in Euclidean space}
\label{App:conv}
The covariantized Pauli matrices in Euclidean space are defined as

\begin{eqnarray}
     & \sigma_\mu&=(-i\vec{\sigma},1) \nonumber\\
    & \bar{\sigma}_\mu&=(i\vec{\sigma},1)
\end{eqnarray}
and
\begin{equation}
\sigma_\mu\bar{\sigma}_\nu+\sigma_\nu\bar{\sigma}_\mu=2\delta_{\mu\nu}
\end{equation}

and the corresponding gamma matrices  are defined as
\begin{align}
    \gamma^\mu&=\begin{pmatrix}
    0 & \sigma^\mu \\
    \bar{\sigma}^\mu & 0 
    \end{pmatrix} & \gamma^5&=\begin{pmatrix}
    -1 & 0 \\
    0 & 1 
    \end{pmatrix}
\end{align}

In $SU(N_c)$ color space, $\vec{\tau}$ is an $N_c\times N_c$ valued matrix with the $2\times2$ Pauli matrices embedded in the upper left corner
\begin{align}
     \tau^+_\mu&=(\vec{\tau},-i)& \tau^-_\mu&=(\vec{\tau},i)
\end{align}
They satisfy the identities
\begin{align}
      \tau^-_\mu\tau^+_\nu-\tau^-_\nu\tau^+_\mu&=2i\bar{\eta}^a_{\mu\nu}\tau^a \\ \tau^+_\mu\tau^-_\nu-\tau^+_\nu\tau^-_\mu&=2i\eta^a_{\mu\nu}\tau^a
\end{align}
where the 't-Hooft symbol is defined in \cite{Liu:2024rdm,Vainshtein:1981wh} 
\begin{equation}
    \eta^{a}_{\mu\nu}=\begin{cases}
        \epsilon^{a}{}_{\mu\nu} ,\ & \mu\neq4,\ \nu\neq4\\
        \delta^{a}_{\mu} ,\ & \mu\neq4,\ \nu=4 \\
        -\delta^{a}_\nu ,\ & \mu=4,\ \nu\neq4
    \end{cases}
\end{equation}
and its  conjugate,
\begin{equation}
    \bar{\eta}^{a}_{\mu\nu}=\begin{cases}
        \epsilon^{a}{}_{\mu\nu} ,\ & \mu\neq4,\ \nu\neq4\\
        -\delta^{a}_\mu ,\ & \mu\neq4,\ \nu=4 \\
        \delta^{a}_\nu ,\ & \mu=4,\ \nu\neq4
    \end{cases}
\end{equation}

The 't Hooft symbol $\eta^a_{\mu\nu}$ satisfies the following identities.

\begin{align*}
&\eta^{a}_{\mu \nu} \eta^{a}_{\rho \lambda} = \delta_{\mu \rho} \delta_{\nu \lambda} - \delta_{\mu \lambda} \delta_{\nu \rho} + \epsilon_{\mu \nu \rho \lambda} \\
&\epsilon_{\mu \nu \rho \lambda} \eta^{a}_{\sigma \lambda} = \delta_{\sigma \mu} \eta^{a}_{\nu \rho} - \delta_{\sigma \nu} \eta^{a}_{\mu \rho} + \delta_{\sigma \rho} \eta^{a}_{\mu \nu} \\
&\eta^{a}_{\mu \nu} \eta^{b}_{\mu \lambda} = \delta_{ab} \delta_{\nu \lambda} + \epsilon_{abc} \eta^{c}_{\nu \lambda} \\
&\epsilon_{abc} \eta^{b}_{\mu \nu} \eta^{c}_{\rho \lambda} = \delta_{\mu \rho} \eta^{a}_{\nu \lambda} - \delta_{\nu \rho} \eta^{a}_{\mu \lambda} - \delta_{\mu \lambda} \eta^{a}_{\nu \rho} + \delta_{\nu \lambda} \eta^{a}_{\mu \rho}
\end{align*}

Same identity applies to $\bar\eta^a_{\mu\nu}$ by replacing 4-d Levi-Civita symbol $\epsilon_{\mu\nu\rho\lambda}\rightarrow-\epsilon_{\mu\nu\rho\lambda}$

\section{Instanton in singular gauge}
\label{App:singular}
The BPST instanton in singular gauge is given by
\begin{equation}
\label{FSTX}
A^a_{\mu}(x;\Omega_I)=R^{ab}(U_I)A^b_\mu(x-z_I)
\end{equation}
which is seen to satisfy both fixed-point and covariant gauge. The profile function is defined as
\begin{equation}
\begin{aligned}
\label{INSINGULAR}
   A^a_\mu(x)=&-\frac{1}{g}\bar\eta^{a}_{\mu\nu}\partial_\nu\ln\Pi(x)=\frac{1}{g}\frac{2\bar\eta^{a}_{\mu\nu}x_\nu\rho^2}{x^2(x^2+\rho^2)}
\end{aligned}
\end{equation}

In  momentum space  it reads
\begin{equation}
\label{INSINGULARQ}
A_\mu^a(q)=i\frac{4\pi^2\rho^2}{g}\frac{\bar\eta^a_{\mu\nu}q_\nu}{q^2}\mathcal{F}_g(\rho q)
\end{equation}
where $\mathcal{F}_g(\rho q)$ is defined in Eq. \eqref{eq:g_form}. 

Here the instanton moduli is captured by $\Omega_I=(z_I,\rho,U_I)$ the rigid color rotation $U_I$, instanton 
location $z_I$ and size $\rho$,  with the singular gauge potential 
\begin{equation}
\label{INSINGULAR}
\Pi(x)=1+\frac{\rho^2}{x^2}
\end{equation}

The rigid color rotation  
$$R^{ab}(U_I)=\frac{1}{2}\mathrm{Tr}(\tau^aU_I\tau^bU_I^\dagger)$$
is defined with $\tau^a$ as an  $N_c\times N_c$ matrix with $2\times2$ Pauli matrices embedded in the upper left corner. For the anti-instanton field, one can substitute  $\bar{\eta}^a_{\mu\nu}$ by $\eta^a_{\mu\nu}$ and flip the sign in front of Levi-Cevita tensor, $\epsilon_{\mu\nu\rho\lambda}\rightarrow-\epsilon_{\mu\nu\rho\lambda}$.

The corresponding field strength profile is defined as
\begin{widetext}
\begin{equation}
    F^a_{\mu\nu}(x)=
        \frac{1}{g}\frac{8\rho^2}{(x^2+\rho^2)^2}\left[\bar{\eta}^a_{\mu\rho}\left(\frac{x_\rho x_\nu}{x^2}-\frac{1}{4}\delta_{\rho\nu}\right)-\bar{\eta}^a_{\nu\rho}\left(\frac{x_\rho x_\mu}{x^2}-\frac{1}{4}\delta_{\rho\mu}\right)\right]
\end{equation}
\end{widetext}

\section{QCD instanton zero modes}
\label{App:ZM}
The quark zero mode solves
\begin{equation}
i\slashed{D}\phi_I(x)=0
\end{equation}
in the instanton background where in fundamental representation, the covariant derivative is defined as
\begin{equation}
    D_\mu=\partial_\mu-i A^a_{\mu}T^a
\end{equation}
where $A^a_{\mu}$ here is the instanton field in singluar gauge.
The solution is left-handed
\begin{equation}
    \phi_I(x)=\varphi(x)\slashed{x}\frac{1-\gamma^5}{2}U\chi
\end{equation}
where $\chi$ is a 4-spinor composed of a $SU(2)$-spinor $\chi_{L,R}$ on both chirality carrying $SU(2)$-color with color-spin locked by $\chi^{i\alpha}_{L,R}=\epsilon^{i\alpha}$, $i$ for spin and $\alpha$ for color. The zero mode profile
in singular gauge is
\begin{equation}
    \varphi(x)=\frac{\rho}{\pi|x|(x^2+\rho^2)^{3/2}}
\end{equation}
In momentum space it is 
\begin{equation}
    \phi_I(k)=-\frac{i\slashed{k}}{k}\varphi^{\prime}(k)\frac{1-\gamma^5}{2}U\chi
\end{equation}
with
\begin{equation}
\label{VARPHIP}
    \varphi^{\prime}(k)=\pi\rho^2\bigg(I_0K_0(z)-I_1K_1(z)\bigg)^\prime_{z=\rho k/2}
\end{equation}
The instanton zero mode is normalized by
\begin{equation}
    \int d^4x \phi^\dagger_{I}(x)\phi_{I}(x)=\int \frac{d^4k}{(2\pi)^4} \phi^\dagger_{I}(k)\phi_{I}(k) =1
\end{equation}
including sum over the spin and color indices.
For the anti-instanton, the zero mode is right handed through the substitution  $\gamma^5\leftrightarrow-\gamma^5$.

The 4-spinor $\chi=\begin{pmatrix}
    \chi_L \\
    \chi_R
\end{pmatrix}$ identity is
\begin{eqnarray}
    &&\chi_L\chi^\dagger_L=\frac{1}{8}\tau_\mu^-\tau_\nu^+\gamma_\mu\gamma_\nu\frac{1-\gamma^5}{2} \nonumber\\[5pt]
    &&\chi_R\chi^\dagger_R=\frac{1}{8}\tau_\mu^+\tau_\nu^-\gamma_\mu\gamma_\nu\frac{1+\gamma^5}{2} \nonumber\\[5pt]
    &&\chi_L\chi^\dagger_R=-\frac{i}{2}\tau_\mu^-\gamma_\mu\frac{1+\gamma^5}{2} \nonumber\\[5pt]
    &&\chi_R\chi^\dagger_L=\frac{i}{2}\tau_\mu^+\gamma_\mu\frac{1-\gamma^5}{2}
\end{eqnarray}
with normalization $(\chi^{i\alpha}_{L,R})^\dagger\chi^{i\alpha}_{L,R}=2$.
\section{Quark propagator in single instanton background}
\label{App:NZM}
Here the quark propagators in a single instanton background is quickly reviewed.

The effects of quark masses on the non-zero mode quark propagator in an instanton or anti-instanton background are not known in closed form \cite{Liu:2021evw}, but for small masses, the quark propagator can be expanded around the chiral limit~\cite{Brown:1978yj}.

\begin{equation}
\begin{aligned}
\label{eq:SIA_prop_q}
    S_I(x,y)=&S_{\mathrm{ZM}}(x,y)+S_0(x-y)\\
    &+\left[S_{\mathrm{NZM}}(x,y)-S_0(x-y)\right]\\
    &-im\Delta_I(x,y)+\mathcal{O}(m^2)
\end{aligned}
\end{equation}
The quark zero mode propagator is defined as

\begin{equation}
\begin{aligned}
S_{\rm ZM}(x,y)=\frac{\phi_I(x)\phi_I^{\dagger}(y)}{im}
\end{aligned}
\end{equation}

The quark propagator with zero modes substracted is defined as
\begin{widetext}
\begin{equation}
\begin{aligned}
i\slashed{D}S_{\rm NZM}(x,y)=\delta^4(x-y)-\phi_I(x)\phi_I^{\dagger}(y)=\frac{1\pm\gamma^5}2\delta^4(x-y) +i\overrightarrow{\slashed{D}}\Delta(x,y) i\overleftarrow{\slashed{D}}\frac{1\mp\gamma^5}2
\end{aligned}
\end{equation}
where the covariant derivative is defined in fundamental representation.

The subtraction of the quark zero mode can be expressed by the isospin-$1/2$ scalar propagator in the instanton background $\Delta(x,y)$. The massless scalar propagator in the single instanton background field is defined as \cite{PhysRevD.17.1583}

\begin{equation}
\begin{aligned}
    \Delta(x,y)=&\frac{1}{4\pi^2(x-y)^2}\left(1+\rho^2\frac{x_\mu y_\nu}{x^2y^2}U\tau_\mu^-\tau_\nu^+U^\dagger\right)\frac{1}{\Pi(x)^{1/2}\Pi(y)^{1/2}}\\
    =&\frac{1}{4\pi^2(x-y)^2}\left(1+\rho^2\frac{x\cdot y}{x^2y^2}+\rho^2\frac{i\bar{\eta}^b_{\mu\nu}x_\mu y_\nu}{x^2y^2}\tau^aR^{ab}(U)\right)\frac{1}{\Pi(x)^{1/2}\Pi(y)^{1/2}}
\end{aligned}
\end{equation}
where the singular gauge potential $\Pi(x)$ is defined in \eqref{INSINGULAR}.

The location of the instanton $z_I$ is set to be zero for simplicity and can be recovered by translational symmetry $x\rightarrow x-z_I$ and $y\rightarrow y-z_I$. Now the non-zero mode propagator for quarks in the chiral-split form reads \cite{PhysRevD.17.1583}
\begin{equation}
\begin{aligned}
\label{eq:NZM}
S_{\rm NZM}(x,y)=i\overrightarrow{\slashed{D}}_x\Delta(x,y)\frac{1+\gamma^5}{2}+\Delta(x,y)i\overleftarrow{\slashed{D}}_{y}\frac{1-\gamma^5}{2}=S_{nz}(x,y)\frac{1+\gamma^5}{2}+\bar{S}_{nz}(x,y)\frac{1-\gamma^5}{2}
\end{aligned}
\end{equation}
where the overhead arrows are defined as $$\Delta(x,y)\overleftarrow{D}_\mu=-\frac{\partial}{\partial y_\mu}\Delta(x,y)-i\Delta(x,y)A_\mu(y)$$ and
$$\overrightarrow{D}_\mu\Delta(x,y)=\frac{\partial}{\partial x_\mu}\Delta(x,y)-iA_\mu(x)\Delta(x,y)$$

After a few steps of algebraic calculation, $S_{nz}$ and $\bar{S}_{nz}$ can be recast in the form~\cite{Schafer:1996wv,Zubkov:1997fn,Liu:2021evw}

\begin{equation}
\begin{aligned}
    S_{nz}(x,y)=&\left[\frac{-i(\slashed{x}-\slashed{y})}{2\pi^2(x-y)^4}\left(1+\rho^2\frac{x_\mu y_\nu}{x^2y^2}U\tau_\mu^-\tau_\nu^+U^\dagger\right)-\frac{\rho^2\gamma_\mu}{4\pi^2}\frac{x_\rho(x-y)_\nu y_\lambda}{(x^2+\rho^2)x^2(x-y)^2y^2}U\tau_\rho^-\tau^+_\mu\tau_\nu^-\tau_\lambda^+U^\dagger\right]\\
    &\times\frac{1}{\Pi(x)^{1/2}\Pi(y)^{1/2}}
\end{aligned}
\end{equation}
and
\begin{equation}
\begin{aligned}
    \bar{S}_{nz}(x,y)=&\left[\frac{-i(\slashed{x}-\slashed{y})}{2\pi^2(x-y)^4}\left(1+\rho^2\frac{x_\mu y_\nu}{x^2y^2}U\tau_\mu^-\tau_\nu^+U^\dagger\right)-\frac{\rho^2\gamma_\mu}{4\pi^2}\frac{x_\rho(x-y)_\nu y_\lambda}{(y^2+\rho^2)x^2(x-y)^2y^2}U\tau_\rho^-\tau^+_\nu\tau_\mu^-\tau_\lambda^+U^\dagger\right]\\
    &\times\frac{1}{\Pi(x)^{1/2}\Pi(y)^{1/2}}
\end{aligned}
\end{equation}

Note that the propagator in the anti-instanton background can be obtained via the substitutions  $\tau^-_\mu\leftrightarrow\tau^+_\mu$, and $\gamma^5\leftrightarrow-\gamma^5$. At short distances, as well as far away from the instanton, the propagator reduces to the free one. At intermediate distances, the propagator is modified due to gluon exchanges with the instanton field \cite{Schafer:1995pz,Kock:2020frx}
\begin{equation}
\begin{aligned}
    S_{\rm NZM}(x,y)|_{x\rightarrow y}&\simeq\frac{-i(\slashed{x}-\slashed{y})}{2\pi^2(x-y)^4}-\frac{i}{16\pi^2}\frac{(x-y)_\mu\gamma_\nu}{(x-y)^2}\gamma^5F_{\mu\nu}(x)
\end{aligned}
\end{equation}

This result is consistent with the OPE of the quark propagator in a general background field.

\section{Reduction scheme for nonzero modes}
\label{App:reduction_large_time}
The on-shell reduction of the Euclidean and massless quark propagator in the instanton background is intricate. In principle, it can be achieved through LSZ reduction in the zero momentum limit. For massless quarks, the LSZ reduction in the zero momentum ($k^2\ll Q^2$) limit reads

\begin{equation}
    \int d^4ye^{-ik\cdot y}S_{nz}(x,y)\slashed{k}\psi_L(k)\simeq\frac{e^{-ik\cdot x}}{(1+\rho^2/x^2)^{1/2}}\left[1+(1-e^{ik\cdot x})\frac{\rho^2}{2x^2}\frac{x_\mu k_\nu}{k\cdot x}U\tau^-_\mu\tau^+_\nu U^\dagger\right]\psi_L(k)
\end{equation}
\begin{equation}
    \int d^4ye^{ik\cdot y}\bar{\psi}_R(k)\slashed{k}\bar{S}_{nz}(y,x)\simeq\frac{e^{ik\cdot x}}{(1+\rho^2/x^2)^{1/2}}\left[1+(1-e^{-ik\cdot x})\frac{\rho^2}{2x^2}\frac{k_\mu x_\nu }{k\cdot x}U\tau^-_\mu\tau^+_\nu U^\dagger\right]\bar{\psi}_R(k)
\end{equation}
\end{widetext}

In the asymptotic limit $x^2\gg \rho^2$, the reduction yields an on-shell free quark.

\section{Averaging over colors}
\label{App:average}

\subsection{Creutz formula}
One way to carry out the color averaging in \eqref{eq:tHooft_g} is by determinantal reduction~\cite{creutz1978invariant}
\begin{equation}
    \int dU\prod_{i=1}^{N_c}U_{a_ib_i}=\frac{1}{N_c!}\epsilon_{a_1\cdots a_{N_c}}\epsilon_{b_1\cdots b_{N_c}}
\end{equation}
and

\begin{equation}
\begin{aligned}
&U^\dagger_{ba}=\frac{1}{(N_c-1)!}\\
&\times\epsilon_{aa_1\cdots a_{N_c-1}}\epsilon_{bb_1\cdots b_{N_c-1}}U_{a_1b_1}\cdots U_{a_{N_c-1}b_{N_c-1}}    
\end{aligned}
\end{equation}
where $\epsilon_{a_1\cdots a_{N_c}}$ is the Levi-Civita tensor of rank-$N_c$ with $\epsilon_{12\cdots N_c}=1$. Now the color averagings of $(UU^\dagger)^p$ are

\begin{enumerate}
    \item $p=1$
      \begin{equation}
      \int dU U_{ab}U^{\dagger}_{cd}=\frac{1}{N_c}\delta_{ad}\delta_{cb}
      \end{equation}
    \item $p=2$
      \begin{equation}
      \begin{aligned}
      &\int dU U_{a_1b_1}U^{\dagger}_{c_1d_1}U_{a_2b_2}U^{\dagger}_{c_2d_2}\\
      =\frac{1}{N_c^2-1}&\big(\delta_{a_1d_1}\delta_{a_2d_2}\delta_{c_1b_1}\delta_{c_2b_2}\\
      &+\delta_{a_1d_2}\delta_{a_2d_1}\delta_{c_1b_2}\delta_{c_2b_1}\big)\\
      -\frac{1}{N_c(N_c^2-1)}&\big(\delta_{a_1d_1}\delta_{a_2d_2}\delta_{c_1b_2}\delta_{c_2b_1}\\
      &+\delta_{a_1d_2}\delta_{a_2d_1}\delta_{c_1b_1}\delta_{c_2b_2}\big)
      \end{aligned}
      \end{equation}
      
\begin{widetext}    
    \item $p=3$
      \begin{equation}
      \begin{aligned}
      &\int dU U_{a_1b_1}U^{\dagger}_{c_1d_1}U_{a_2b_2}U^{\dagger}_{c_2d_2}U_{a_3b_3}U^{\dagger}_{c_3d_3}
      =\frac{N_c^2-2}{N_c(N^2_c-4)(N_c^2-1)}\\
      &\times(\delta_{a_1d_1}\delta_{a_2d_2}\delta_{a_3d_3}\delta_{c_1b_1}\delta_{c_2b_2}\delta_{c_3b_3}
      +\delta_{a_1d_2}\delta_{a_2d_1}\delta_{a_3d_3}\delta_{c_1b_2}\delta_{c_2b_1}\delta_{c_3b_3}
      +\delta_{a_1d_3}\delta_{a_2d_2}\delta_{a_3d_1}\delta_{c_1b_3}\delta_{c_2b_2}\delta_{c_3b_1}\\
      &+\delta_{a_1d_1}\delta_{a_3d_2}\delta_{a_2d_3}\delta_{c_1b_1}\delta_{c_3b_2}\delta_{c_2b_3}
      +\delta_{a_1d_3}\delta_{a_3d_2}\delta_{a_2d_1}\delta_{c_1b_3}\delta_{c_3b_2}\delta_{c_2b_1}
      +\delta_{a_1d_2}\delta_{a_2d_3}\delta_{a_3d_1}\delta_{c_1b_2}\delta_{c_2b_3}\delta_{c_3b_1})\\
      &-\frac{1}{(N_c^2-4)(N_c^2-1)}\\
      &\times(\delta_{a_1d_1}\delta_{a_2d_2}\delta_{a_3d_3}\delta_{c_1b_2}\delta_{c_2b_1}\delta_{c_3b_3}
      +\delta_{a_1d_2}\delta_{a_2d_1}\delta_{a_3d_3}\delta_{c_1b_1}\delta_{c_2b_2}\delta_{c_3b_3}
      +\delta_{a_1d_1}\delta_{a_2d_2}\delta_{a_3d_3}\delta_{c_1b_3}\delta_{c_2b_2}\delta_{c_3b_1}\\
      &+\delta_{a_1d_3}\delta_{a_2d_2}\delta_{a_3d_1}\delta_{c_1b_1}\delta_{c_2b_2}\delta_{c_3b_3}
      +\delta_{a_1d_1}\delta_{a_2d_2}\delta_{a_3d_3}\delta_{c_1b_1}\delta_{c_3b_2}\delta_{c_2b_3}
      +\delta_{a_1d_1}\delta_{a_3d_2}\delta_{a_2d_3}\delta_{c_1b_1}\delta_{c_2b_2}\delta_{c_3b_3}\\
      &+\delta_{a_1d_3}\delta_{a_3d_2}\delta_{a_2d_1}\delta_{c_1b_1}\delta_{c_3b_2}\delta_{c_2b_3}
      +\delta_{a_1d_3}\delta_{a_3d_2}\delta_{a_2d_1}\delta_{c_3b_1}\delta_{c_2b_2}\delta_{c_1b_3}
      +\delta_{a_1d_3}\delta_{a_3d_2}\delta_{a_2d_1}\delta_{c_1b_2}\delta_{c_2b_1}\delta_{c_3b_3}\\
      &+\delta_{a_1d_1}\delta_{a_3d_2}\delta_{a_2d_3}\delta_{c_1b_3}\delta_{c_3b_2}\delta_{c_2b_1}
      +\delta_{a_1d_1}\delta_{a_3d_2}\delta_{a_2d_3}\delta_{c_1b_3}\delta_{c_3b_2}\delta_{c_2b_1}
      +\delta_{a_1d_1}\delta_{a_3d_2}\delta_{a_2d_3}\delta_{c_1b_3}\delta_{c_3b_2}\delta_{c_2b_1}\\
      &+\delta_{a_1d_2}\delta_{a_2d_3}\delta_{a_3d_1}\delta_{c_1b_1}\delta_{c_3b_2}\delta_{c_2b_3}
      +\delta_{a_1d_2}\delta_{a_2d_3}\delta_{a_3d_1}\delta_{c_3b_1}\delta_{c_2b_2}\delta_{c_1b_3}
      +\delta_{a_1d_2}\delta_{a_2d_3}\delta_{a_3d_1}\delta_{c_1b_2}\delta_{c_2b_1}\delta_{c_3b_3}\\
      &+\delta_{a_1d_1}\delta_{a_3d_2}\delta_{a_2d_3}\delta_{c_1b_2}\delta_{c_2b_3}\delta_{c_3b_1}
      +\delta_{a_1d_1}\delta_{a_3d_2}\delta_{a_2d_3}\delta_{c_1b_2}\delta_{c_2b_3}\delta_{c_3b_1}
      +\delta_{a_1d_1}\delta_{a_3d_2}\delta_{a_2d_3}\delta_{c_1b_2}\delta_{c_2b_3}\delta_{c_3b_1})\\
      &+\frac{2}{N_c(N_c^2-4)(N_c^2-1)}\\
      &\times(\delta_{a_1d_2}\delta_{a_2d_3}\delta_{a_3d_1}\delta_{c_1b_1}\delta_{c_2b_2}\delta_{c_3b_3}
      +\delta_{a_1d_1}\delta_{a_2d_2}\delta_{a_3d_3}\delta_{c_1b_2}\delta_{c_2b_3}\delta_{c_3b_1}
      +\delta_{a_1d_3}\delta_{a_3d_2}\delta_{a_2d_1}\delta_{c_1b_1}\delta_{c_2b_2}\delta_{c_3b_3}\\
      &+\delta_{a_1d_1}\delta_{a_2d_2}\delta_{a_3d_3}\delta_{c_1b_3}\delta_{c_3b_2}\delta_{c_2b_1}
      +\delta_{a_1d_2}\delta_{a_2d_3}\delta_{a_3d_1}\delta_{c_1b_3}\delta_{c_3b_2}\delta_{c_2b_1}
      +\delta_{a_1d_3}\delta_{a_3d_2}\delta_{a_2d_1}\delta_{c_1b_2}\delta_{c_2b_3}\delta_{c_3b_1}\\
      &+\delta_{a_1d_1}\delta_{a_3d_2}\delta_{a_2d_3}\delta_{c_3b_2}\delta_{c_2b_2}\delta_{c_2b_3}
      +\delta_{a_1d_1}\delta_{a_3d_2}\delta_{a_2d_3}\delta_{c_1b_2}\delta_{c_2b_1}\delta_{c_3b_3}
      +\delta_{a_1d_3}\delta_{a_2d_2}\delta_{a_3d_1}\delta_{c_1b_2}\delta_{c_2b_1}\delta_{c_3b_3}\\
      &+\delta_{a_1d_3}\delta_{a_2d_2}\delta_{a_3d_1}\delta_{c_1b_1}\delta_{c_3b_2}\delta_{c_2b_3}
      +\delta_{a_1d_2}\delta_{a_2d_1}\delta_{a_3d_3}\delta_{c_1b_1}\delta_{c_3b_2}\delta_{c_2b_3}
      +\delta_{a_1d_2}\delta_{a_2d_1}\delta_{a_3d_3}\delta_{c_3b_2}\delta_{c_2b_2}\delta_{c_2b_3})
      \end{aligned}
      \end{equation}
\end{widetext}  
\end{enumerate}

\subsection{CNZ formula}

\cite{Chernyshev:1995gj}
However, For large values of $p$, this averaging method is tedious.  Since $N_c \otimes N_c = 1 \oplus (N_c^2-1)$, the group integral practically reduces to finding all projections of the product of adjoint representations onto the singlet for $SU(N_c)$. The result can be obtained by using the graphical color projection rules~\cite{Chernyshev:1995gj,Nowak:1988bh,Miesch:2023hjt}, 
with the following results
\begin{widetext}
\begin{enumerate}
    \item $p=2$
      \begin{equation}
      \begin{aligned}
      &\int dU U_{a_1b_1}U^{\dagger}_{c_1d_1}U_{a_2b_2}U^{\dagger}_{c_2d_2}=\frac{1}{N_c^2}\delta_{a_1d_1}\delta_{a_2d_2}\delta_{c_1b_1}\delta_{c_2b_2}+\frac{1}{4(N_c^2-1)}\lambda^\alpha_{a_1d_1}\lambda^\alpha_{a_2d_2}\lambda^\beta_{c_1b_1}\lambda^\beta_{c_2b_2}
      \end{aligned}
      \end{equation}

    \item $p=3$
      \begin{equation}
      \begin{aligned}
      &\int dU U_{a_1b_1}U^{\dagger}_{c_1d_1}U_{a_2b_2}U^{\dagger}_{c_2d_2}U_{a_3b_3}U^{\dagger}_{c_3d_3}=\frac{1}{N_c^3}\delta_{a_1d_1}\delta_{a_2d_2}\delta_{a_3d_3}\delta_{c_1b_1}\delta_{c_2b_2}\delta_{c_3b_3}\\
      &+\frac{1}{4N_c(N_c^2-1)}\left(\lambda^\alpha_{a_1d_1}\lambda^\alpha_{a_2d_2}\delta_{a_3d_3}\lambda^\beta_{c_1b_2}\lambda^\beta_{c_2b_1}\delta_{c_3b_3}+\delta_{a_1d_1}\lambda^\alpha_{a_2d_2}\lambda^\alpha_{a_3d_3}\delta_{c_1b_1}\lambda^\beta_{c_2b_2}\lambda^\beta_{c_3b_3}+\lambda^\alpha_{a_1d_1}\delta_{a_2d_2}\lambda^\alpha_{a_3d_3}\lambda^\beta_{c_1b_1}\delta_{c_2b_2}\lambda^\beta_{c_3b_3}\right)\\
       &+\frac{1}{4(N_c^2-1)}\left(\frac{N_c}{2(N_c^2-4)}d^{\alpha\beta\gamma}d^{\alpha'\beta'\gamma'}\lambda^\alpha_{a_1d_1}\lambda^\beta_{a_2d_2}\lambda^\gamma_{a_3d_3}\lambda^{\alpha'}_{c_1b_2}\lambda^{\beta'}_{c_2b_1}\lambda^{\gamma'}_{c_3b_3}\right)\\
       &+\frac{1}{4(N_c^2-1)}\left(\frac{1}{2N_c}f^{\alpha\beta\gamma}f^{\alpha'\beta'\gamma'}\lambda^\alpha_{a_1d_1}\lambda^\beta_{a_2d_2}\lambda^\gamma_{a_3d_3}\lambda^{\alpha'}_{c_1b_2}\lambda^{\beta'}_{c_2b_1}\lambda^{\gamma'}_{c_3b_3}\right)
      \end{aligned}
      \end{equation}
    \item $p=4$
      \begin{equation}
      \begin{aligned}
      &\int dU U_{a_1b_1}U^{\dagger}_{c_1d_1}U_{a_2b_2}U^{\dagger}_{c_2d_2}U_{a_3b_3}U^{\dagger}_{c_3d_3}U_{a_4b_4}U^{\dagger}_{c_4d_4}=\frac{1}{N_c^4}\delta_{a_1d_1}\delta_{a_2d_2}\delta_{a_3d_3}\delta_{a_4d_4}\delta_{c_1b_1}\delta_{c_2b_2}\delta_{c_3b_3}\delta_{c_4b_4}\\
      &+\left[\frac{1}{N_c}\delta_{a_4d_4}\delta_{c_4b_4}\left(\int dU U_{a_1b_1}U^{\dagger}_{c_1d_1}U_{a_2b_2}U^{\dagger}_{c_2d_2}U_{a_3b_3}U^{\dagger}_{c_3d_3}-\frac{1}{N_c^3}\delta_{a_1d_1}\delta_{a_2d_2}\delta_{a_3d_3}\delta_{c_1b_1}\delta_{c_2b_2}\delta_{c_3b_3}\right)+\mathrm{permutations}\right]\\
      &+\lambda^\alpha_{a_1d_1}\lambda^\beta_{a_2d_2}\lambda^\gamma_{a_3d_3}\lambda^\delta_{a_4d_4}\lambda^{\alpha'}_{c_1b_1}\lambda^{\beta'}_{c_2b_2}\lambda^{\gamma'}_{c_3b_3}\lambda^{\delta'}_{c_4b_4}\bigg[\frac{1}{16(N_c^2-1)^2}\left(\delta^{\alpha\beta}\delta^{\gamma\delta}\delta^{\alpha'\beta'}\delta^{\gamma'\delta'}+\delta^{\alpha\gamma}\delta^{\beta\delta}\delta^{\alpha'\gamma'}\delta^{\beta'\delta'}+\delta^{\alpha\delta}\delta^{\beta\gamma}\delta^{\alpha'\delta'}\delta^{\beta'\gamma'}\right)\\
       &+\frac{1}{4(N_c^2-1)}\bigg(\frac{N_c^2}{4(N_c^2-4)^2}d^{\alpha\beta\epsilon}d^{\gamma\delta\epsilon}d^{\alpha'\beta'\rho}d^{\gamma'\delta'\rho}+\frac{1}{4(N_c^2-4)}d^{\alpha\beta\epsilon}f^{\gamma\delta\epsilon}d^{\alpha'\beta'\epsilon'}f^{\gamma'\delta'\epsilon'}\\
       &+\frac{1}{4(N_c^2-4)}f^{\alpha\beta\epsilon}d^{\gamma\delta\epsilon}f^{\alpha'\beta'\epsilon'}d^{\gamma'\delta'\epsilon'}+\frac{1}{4N_c^2}f^{\alpha\beta\epsilon}f^{\gamma\delta\epsilon}f^{\alpha'\beta'\epsilon'}f^{\gamma'\delta'\epsilon}\bigg)\bigg]
      \end{aligned}
      \end{equation}

\end{enumerate}
\end{widetext}

\bibliography{ILM}

\end{document}